\documentclass[%
 aip,
 amsmath,amssymb,
 reprint,%
]{revtex4-1}

\usepackage{graphicx}
\usepackage{dcolumn}
\usepackage{bm}

\usepackage[utf8]{inputenc}
\usepackage[T1]{fontenc}
\usepackage{mathptmx}
\usepackage{etoolbox}
\usepackage{hyperref}

\makeatletter
\def\@email#1#2{%
 \endgroup
 \patchcmd{\titleblock@produce}
  {\frontmatter@RRAPformat}
  {\frontmatter@RRAPformat{\produce@RRAP{*#1\href{mailto:#2}{#2}}}\frontmatter@RRAPformat}
  {}{}
}%
\makeatother
\begin{document}

\preprint{AIP/123-QED}

\title{Kinematic Model of Magnetic Domain Wall Motion for Fast, High-Accuracy Simulations}
\author{Kristi Doleh}
\altaffiliation{first authors contributed equally.}
\author{Leonard Humphrey}
\altaffiliation{first authors contributed equally.}
\author{Chandler M. Linseisen}
\affiliation{Electrical and Computer Engineering, The University of Texas at Dallas, Richardson, TX}
\author{Michael D. Kitcher}
\affiliation{Department of Materials Science \& Engineering, Carnegie Mellon University, Pittsburgh, PA}
\author{Joanna M. Martin}
\affiliation{Electrical and Computer Engineering, The University of Texas at Dallas, Richardson, TX}
\author{Can Cui}
\author{Jean Anne C. Incorvia}
\affiliation{Electrical and Computer Engineering, The University of Texas at Austin, Austin, TX}
\author{Felipe Garcia-Sanchez}
\affiliation{Departamento de Fisica Aplicada, Universidad de Salamanca, Salamanca, ES}
\author{Naimul Hassan}
\author{Alexander J. Edwards}
 
 \author{Joseph S. Friedman}
 
 \email{fgs@usal.es}
 \email{Joseph.Friedman@utdallas.edu}
 \email{Alexander.Edwards@utdallas.edu}
\affiliation{Electrical and Computer Engineering, The University of Texas at Dallas, Richardson, TX}

\date{\today}

\begin{abstract}
Domain wall (DW) devices have garnered recent interest for diverse applications including memory, logic, and neuromorphic primitives; fast, accurate device models are therefore imperative for large-scale system design and verification.  Extant DW motion models are sub-optimal for large-scale system design either over-consuming compute resources with physics-heavy equations or oversimplifying the physics, drastically reducing model accuracy.  We propose a DW model inspired by the phenomenological similarities between motions of a DW and a classical object being acted on by forces like air resistance or static friction.  Our proposed phenomenological model predicts DW motion within 1.2\% on average compared with micromagnetic simulations that are 400 times slower.  Additionally our model is seven times faster than extant collective coordinate models and 14 times more accurate than extant hyper-reduced models making it an essential tool for large-scale DW circuit design and simulation.  The model is publicly posted along with scripts that automatically extract model parameters from user-provided simulation or experimental data to extend the model to alternative micromagnetic parameters.
\end{abstract}

\maketitle


Spintronic technologies  have gained recent interest for use in memory and neuromorphic networks \cite{NeuromorphicSpintronics}.  Specifically, domain wall (DW) devices have been proposed for memory \cite{DW-Racetrack, DWMTJ-IMC} and logic \cite{8715734, U_Patrick_AdderBench, DWMTJ-IMC}, as well as both synaptic \cite{DW-MTJ-Synapse, DW-MTJ-Synapse-Spiking, U_Akinola_2019_STDP} and neuronal \cite{U_MagneticDomainWallNeuron, U_JxCDCAnisotropyNeuron, U_Akinola_2019_STDP, U_Cui_2020_LatInhibit, U_ShapeAnis, KRoy-DW-MTJ-LeakyNeuron} functions in level-based \cite{DW-MTJ-Synapse} and spiking neural networks \cite{U_MagneticDomainWallNeuron, U_JxCDCAnisotropyNeuron, U_Akinola_2019_STDP, DW-MTJ-Synapse-Spiking, U_Cui_2020_LatInhibit, U_ShapeAnis, KRoy-DW-MTJ-LeakyNeuron}. These proposals generally leverage the ability to move the DW through a spin-transfer torque (STT) or spin-orbit torque (SOT) current in a non-volatile manner and often electrically read the DW position using a magnetic tunnel junction (MTJ); the complete DW-MTJ can be treated as an electrical device and integrated directly into analog or digital circuits.


\begin{figure}
    \centering
    \includegraphics[width=0.48\textwidth]{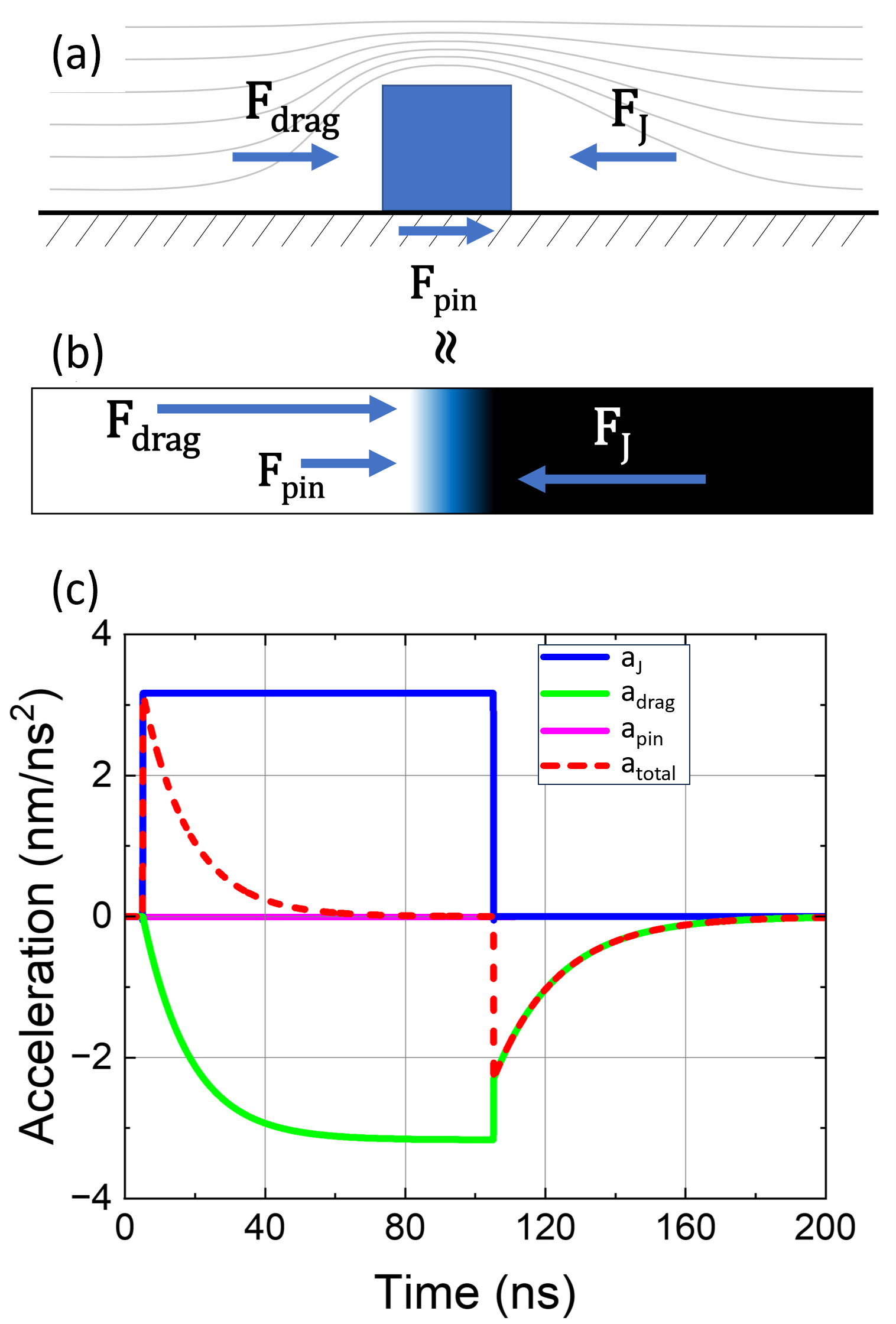}
    \caption{(a) and (b) Free body diagram of DW motion for the proposed model alongside analogous classical system in which a box is pushed along a surface by a constant applied force through a viscous fluid.  In this analogy, the current-induced force $F_J$ is represented by the applied classical force, the micromagnetic damping force is represented by the fluid drag, and the DW pinning force is represented by the static friction when the box is resting. (c) The component accelerations of Equation \ref{eq:acc} applied to the DW when a current pulse is applied and later removed.  As the current density is sufficiently large, there is no pinning.}
    \vspace{-1em}
    \label{fig:apl-fig1}
\end{figure}

Proper design and verification of useful memory, logic, and neuromorphic network systems requires large-scale simulations. As simulating the full micromagnetic behavior of a large system comprising DW devices would require impractically large computational effort, compact device models with simplified physics are required.  This can be challenging, as over-simplification of a device can lead to inaccurate simulation results, whereas overly complex models can contribute to exceedingly large simulation times.

Some previous DW models have been too simplistic, failing to account for some of the more non-linear aspects of DW motion \cite{doi:10.1063/1.3536793, 9180675}.  While these models may be useful for proof-of-concept simulations of circuits and systems, they are insufficiently accurate for designing and optimizing circuits that will be implemented with real devices.

Other previous models, particularly those derived directly from equations like the Landau-Lifshitz-Gilbert (LLG) equation, have been overly complex.  Very-high-accuracy DW models require a hybrid process that incorporates both a micromagnetic solver and a circuit simulator, requiring enormous computational resources \cite{incorvia12, 7430277, 6131575}.  Others have proposed the use of 1D or 2D collective coordinate (CC) models \cite{9662302, Thiaville2006, doi:10.1063/1.1688673, doi:10.1063/1.1667804, THIAVILLE20021061, PhysRevB.86.054445, doi:10.1063/1.4881778, 9072283, PhysRevB.83.245211, NASSERI201825}, which, while accounting for many of the observed phenomena, require more compute resources than necessary to accurately simulate.  Furthermore, these models often do not match experimental data, relying so heavily on idealized equations that they become disconnected from practical devices.  There thus does not exist a model that is both sufficiently fast and accurate for large-scale DW circuit simulations.

Critically, models derived from the LLG equation fail to take advantage of the strong similarities between the motion of a DW and a classical object, illustrated in Fig. \ref{fig:apl-fig1}.  That is, one-dimensional DW motion resembles the kinematic motion of a classical object: both exhibit momentum, terminal velocity, and damping.  By matching observed DW phenomena to their classical counterparts, DW acceleration can be deconstructed into component forces similar to an application of Newton's second law of motion.

Exploiting the similarities between the motion of a DW and classical object, we therefore propose the first DW model that accounts for observed non-linear phenomena while enabling extremely fast computation for large-scale simulation. We develop a low order kinematic model inspired by Stokes' drag equation:

\begin{equation}
    \frac{d^2x}{dt^2} = -A*\frac{dx}{dt} + B
    \label{eq:drag}
\end{equation}

\noindent where the acceleration of the DW is a linear function of the velocity and the external forces, B.
By fitting our low-order kinematic model to highly-accurate micromagnetic simulations, we achieve an ultra-low-cost yet sufficiently accurate model for simulating large DW networks that faithfully predicts the behavior of large-scale systems comprised of DW devices.  

The model is verified against high-precision micromagnetic simulations across a wide range of micromagnetic parameters, faithfully predicting the DW position to within 1.2\% on average.  The Verilog-A domain wall model is provided in the Supplementary Information and posted publicly at \href{https://github.com/AJEdwards314/KinematicDomainWallModel}{Github} and \href{https://personal.utdallas.edu/~joseph.friedman/Papers/KinematicDWModel_05_24.zip}{our website} along with automated scripts for extracting model parameters from user-provided micromagnetic simulation or experimental results, enabling effortless extension of the Verilog-A model to alternative micromagnetic parameters or stimuli.

In our proposed DW model, the DW is treated as an object in 1-dimensional space with continuous and differentiable position and velocity. A counterpart to Newton's second law can therefore be applied, expressing DW motion in terms of forces and resulting in a second-order ordinary differential equation for the DW velocity:


\begin{equation}
        x''(t) = v'(t) = a(t) =  a_J(t) + a_{damp}(t) + a_{pin}(t),
\end{equation}
\noindent with
\begin{equation}
        a_J(t) = \frac{F_J}{m},\; a_{damp} = \frac{F_{damp}}{m},\; a_{pin} = \frac{F_{pin}}{m},
        \label{eq:acc}
\end{equation}
where $x$ is the DW position, $v$ is the DW velocity, $a$ is the DW acceleration, $m$ is the DW effective mass, $a_J$ ($F_J$) is current-induced acceleration (force), $a_{drag}$ ($F_{drag}$) is the damping acceleration (force), and $a_{pin}$ ($F_{pin}$) is the pinning acceleration (force) when the DW is not moving. Figure \ref{fig:apl-fig1} illustrates the proposed DW model alongside an analogous classical system.  An argument from first principles justifying this second-order kinematic reduction is detailed in the Supplementary Information.

The component forces are based on classical counterparts and are modeled with polynomial fitting equations so as to closely match micromagnetic simulations. These fitting polynomials incorporate a number of micromagnetic parameters including effective mass, and are chosen to minimize model error relative to micromagnetic simulations.  The model is trained on a range of micromagnetic and geometric parameter corners, leaving DW position as a hysteretic function of only current density and time.  The model is reproduced in the Supplementary Information and posted publicly online along with scripts that automatically extract model fitting parameters from user-provided simulation or experimental data to extend the model beyond the 32 micromagnetic parameter corners evaluated here.  The use of polynomial fitting functions as compared with the copious trigonometric evaluations required by conventional CC models provides our proposed model with a significant speedup with minimal loss in accuracy as demonstrated later in this work. 



DW motion is stimulated using electrical current, applying a constant force to the DW.  When a current is applied to the heavy metal layer of a DW device, DW motion is induced due to various micromagnetic effects including, but not limited to, SOT.  This effective force -- described by $a_J$ -- is a function of only the current density, $J$, and is therefore approximated with a quartic polynomial of $J$:

\begin{equation}
a_J = k_4J^4 + k_3J^3 + k_2J^2 + k_1J + k_0, \quad J \geq 0,
\end{equation}

\noindent where $k_0$ - $k_4$ are fitting parameters.  For $J < 0$, $a_J(J) = -a_J(-J)$, to ensure equal and opposite behavior for positive and negative current densities.

While this model has not been developed to incorporate STT or magnetic field stimuli, this kinematic DW model can be directly applied to such situations by solving for the appropriate model fitting terms with the approach described later in this work.


The damping force experienced by a DW limits the DW to a maximum velocity, as with an object moving through a viscous fluid.  We model the damping as Stokes' drag: a force proportional to and opposing the velocity.  As the magnitude of the damping changes with applied current density \cite{Torrejon2016}, a linear function of $J$ is incorporated in the model to account for this effect:

\begin{equation}
        a_{damp} = -v * (d_1 + d_2|J|),
\end{equation}

\noindent where $d_1$ and $d_2$ are fitting parameters.  Under a constant applied current density, the DW accelerates until it reaches a terminal velocity when $a_J + a_{damp} = 0$.  For a given micromagnetic parameter corner, this terminal velocity may be defined in terms of current density allowing $k_0$, $\cdots$, $k_4$, $d_1$, and $d_2$ to be determined simultaneously to closely match terminal velocity and acceleration time-constants extracted from the micromagnetic simulations.


DW pinning is modeled like static friction, such that if the DW is not moving, the pinning force will counter motion until the applied force is greater than a threshold.  The force due to pinning is described by the following piecewise function:

\begin{equation}
F_{pin} = 
\left\{
\begin{array}{cl}
     -F_J, & \vert J\vert < p_1 \:\&\: \vert v\vert < p_2\\
     0, & otherwise
\end{array}
\right.
\end{equation}

\noindent where the thresholds $p_1$ and $p_2$ are tunable model parameters.


As the proposed kinematic model is 1-dimensional, track width is incorporated into the fitting parameters.  Track length is a tunable parameter, and the DW is modeled to bounce off the track ends according to 

\begin{equation}
    v_\text{post\_bounce} = - c_r * v_\text{pre\_bounce}, 
\end{equation}

\noindent where the coefficient of restitution $c_r \in [0,1]$ is a tunable model parameter.

Our kinematic model is demonstrated to be ultra-efficient with reasonably high accuracy compared with extant DW models.  Micromagnetic simulations were used to glean model fitting parameters and benchmark accuracy.  Our compact model was implemented in SPICE and simulated, demonstrating minimal deviation from micromagnetic simulation and significant speedup relative to hybrid micromagnetic models and CC models.


High-accuracy micromagnetic simulations were performed using mumax3 to develop and evaluate the model.  As highlighted in Table \ref{tab:parameters}, DW track simulations were run across a large range of parameter corners to increase the application space of the model.  These parameters include track width, $M_{sat}, B_{anis,eff},\alpha,$and $Aex$, with $Ku$ determined from $M_{sat}$ and $B_{anis,eff}$ (further details are in the Supplementary Information).

Numerous training simulations were run across a dense range of current pulse densities $J$ and pulse duration $\tau$.  The micromagnetic simulations comprised a DW in an infinitely long track stimulated by a single square current pulse of density $J$ and duration $\tau$.  For each parameter corner, simulations were run over a range of ten current densities for a single pulse length, which was sufficiently long enough to perform model fitting for each corner. 

Model parameters were chosen so as to minimize model error relative to DW position and velocity extracted from the micromagnetic simulations.  For each micromagnetic parameter corner, a set of model fitting parameters is chosen to correctly predict DW motion for all of the simulated pulses with varying $J$ and $\tau$ run on that parameter corner, and this process is detailed in the Supplementary Information.  Therefore, for a single parameter corner, the model will be able to properly respond to a wide range of current pulses, including pulses of strange shapes. Through this process, we can therefore ensure our model remains accurate over a reasonable set of parameter combinations and input pulses.  Additionally, the model may easily be extended to alternative micromagnetic parameters or stimuli, as described in the following paragraphs.
    

\begin{table}[t]
\caption{Micromagnetic parameters}
\centering
\begin{tabular}{ | m{ 0.24\textwidth} | m{ 0.2\textwidth} | }
\hline
\textbf{Parameter} & \textbf{Value(s)} \\
\hline
 Width [nm] & 50, 100\\ 
 \hline
 Length [nm] & 500\\  
 \hline
 Thickness [nm] & 1.2\\ 
 \hline
 Current Density (J) [mA/$\mu$ m$^2$] & $8$ to $800$\\ 
 \hline
 Saturation Magnetization (Msat) [A/m] & $7.95 $x$ 10^5$, $1.2 $x$ 10^6$ \\ 
 \hline
 $B_{anis,eff}$ [mT] & 20, 350 \\ 
 \hline
 Anisotropy (Ku) [J/m$^3$] & $4.05 $x$ 10^5$, $5.36$x$ 10^5$,$9.17 $x$ 10^5$, $1.11 $x$ 10^6$ \\ 
 \hline
 Gilbert Damping Factor ($\alpha$) & 0.01, 0.05 \\ 
 \hline
 Exchange stiffness (Aex) [J/m] & $11 $x$ 10^{-12}$, $31 $x$ 10^{-12}$ \\
 \hline
 Current Pulse Width [ns] & $20$ to $100$ \\
 \hline
\end{tabular}
 \label{tab:parameters}
\end{table}

%

\begin{figure}
    \centering
    \includegraphics[width=0.48\textwidth]{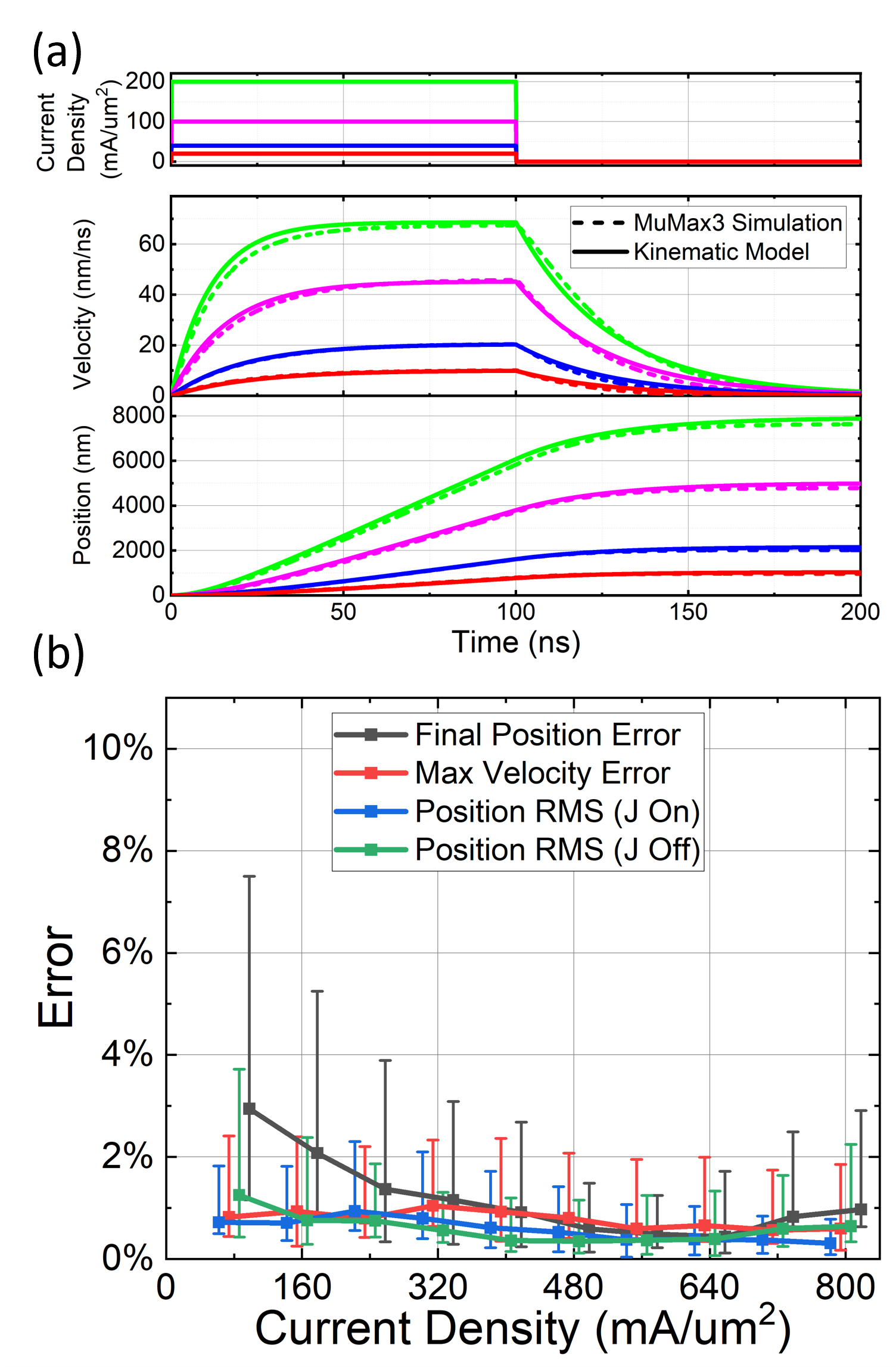}
    \caption{(a) DW position in response to a current pulse, for both micromagnetic simulation and the kinematic model, with Aex = 11x$10^{-12}$ J/m, Ku = 4.05x$10^5$ J/m$^3$, $\alpha$ = 0.01, $M_{\text{sat}}$ = 7.95x$10^5$ A/m, and track width = 100 nm. (b) Median, lower quartile, and upper quartile of four different error metrics averaged across tested values of all parameters.}
    \vspace{-1em}
    \label{fig:apl-fig2}
\end{figure}

We have implemented the proposed kinematic DW model in Verilog-A, thereby enabling direct SPICE simulations of large-scale systems; this model is included in the Supplementary Information and posted publicly at \href{https://github.com/AJEdwards314/KinematicDomainWallModel}{Github} and \href{https://personal.utdallas.edu/~joseph.friedman/Papers/KinematicDWModel_05_24.zip}{our website}.  Fitting parameters for the various parameter corners are left as model parameters that can be programmed by the user, and our table of inferred model fitting terms for each micromagnetic parameter corner is included in the Supplementary Information.  While fitting parameters for parameter combinations not listed in the table may be interpolated from the existing data, future users should be aware that the model has only been verified against micromagnetic simulations at the listed parameter corners.  

We have also supplied the scripts necessary should the reader wish to extend the model to a larger parameter space or alternate stimuli such as STT or magnetic field.  The template mumax3 simulation script that is reproduced in the Supplementary Information and posted on \href{https://github.com/AJEdwards314/KinematicDomainWallModel}{Github} and \href{https://personal.utdallas.edu/~joseph.friedman/Papers/KinematicDWModel_05_24.zip}{our website} can be altered to match the desired parameter space or stimuli.  Data should be collected across a range of stimuli strengths and pulse widths, and model fitting parameters can be automatically extracted using the Matlab scripts provided in the Supplementary Information and posted on \href{https://github.com/AJEdwards314/KinematicDomainWallModel}{Github} and \href{https://personal.utdallas.edu/~joseph.friedman/Papers/KinematicDWModel_05_24.zip}{our website}; the presented model is therefore flexible, with built-in mechanisms to assimilate new simulation or experimental data.

The kinematic DW model is directly amenable to DW-MTJ circuit simulations, though it is not limited to these applications; when used as a DW-MTJ, the modeled electrical behavior of the tunnel barrier is consistent with existing DW-MTJ models in the literature: it is modeled as a pair of parallel resistors, $R_P$ and $R_{AP}$, according to:

\begin{equation}
    R_{DW-MTJ} = \frac{R_P}{x} \bigg\vert \bigg\vert \frac{R_{AP}}{1-x},
\end{equation}

\noindent where $x$ is the position of the DW normalized to the length of the track and $R_P$ ($R_{AP}$) is the (anti-)parallel resistance of the device when $x = 1$($0$) \cite{8715734}.  Track length, $R_P$, and $R_{AP}$ are left as model parameters for the user.

%

\begin{figure*}[t]
    \centering
    \includegraphics[width=\textwidth]{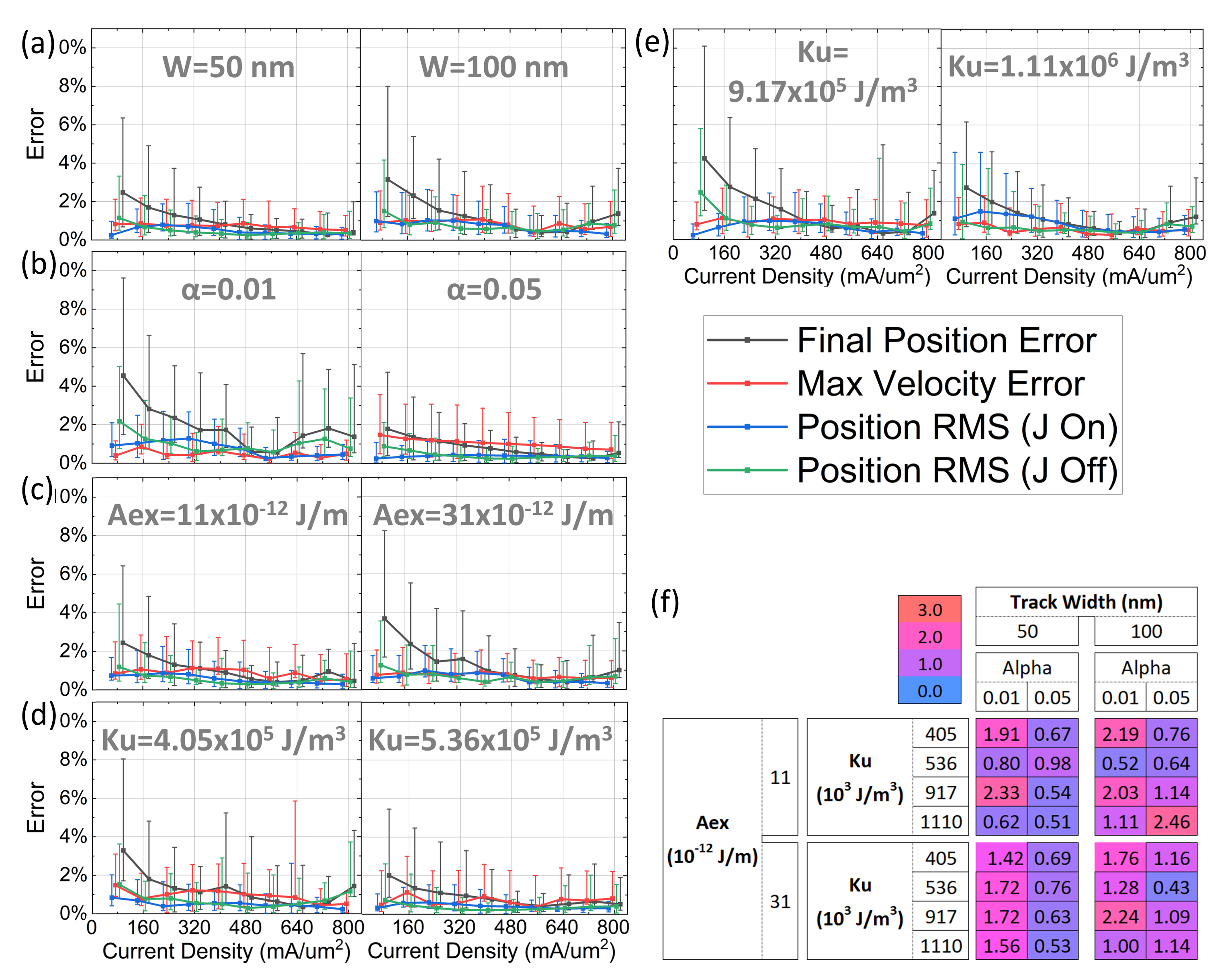}
    \caption{(a) - (e) Median, lower quartile, and upper quartile of four different error metrics averaged across tested values of all parameters but one: $\alpha$ in (a), W in (b), Aex in (c), and Ku in (d) and (e). (f) Averaged percent error over the ten runs of various current density $J$ for each parameter corner simulated.  In total this figure summarizes our model performance over 320 micromagnetic simulations.}
    \vspace{-1em}
    \label{fig:apl-fig3}
\end{figure*}


For a single current pulse, the Verilog-A model matches very closely with the micromagnetic results across all parameter corners. Fig. \ref{fig:apl-fig2}(a) plots the DW position and velocity over time in response to various current densities for one of these parameter corners, highlighting the similarities between the kinematic model and mumax3 simulation.

The accuracy of the model was determined to be reasonably high across all reported parameter corners.  Over each parameter corner, the accuracy of the model was analyzed using various error metrics including error in the final DW displacement, in the maximum velocity reached during the simulation, and the root mean square (RMS) error of the DW position over time. The error in the final DW position is most indicative of model performance for simulations with constant pulse width $\tau$, whereas the error in maximum velocity is more indicative of model error when $\tau$ changes or is notably long, \textit{i.e.} cases where the transient error would average out or become negligible. The RMS error in DW position was evaluated during the two transient periods of the simulation, the first being the time it took to reach maximum velocity and the second being the time it took to stop.

As illustrated in Figs. \ref{fig:apl-fig2}(b) and \ref{fig:apl-fig3}(a)-(e), the model demonstrates low error, $<1.2\%$ on average.  Fig. \ref{fig:apl-fig3}(a)-(e) depict the error averaged across the various values of all parameters except one which is held constant according to the figure caption.  The model performs well in every case, though there are some parameter values where the model is more accurate; for instance, Fig. \ref{fig:apl-fig3}(b) illustrates that the model is more accurate with higher Gilbert damping.

Additionally, Fig. \ref{fig:apl-fig3}(f) shows high model performance at each parameter corner.  For each corner, simulations were run with a range of ten values for $J$ and a sufficiently large $\tau$ for the DW to reach its maximum velocity.  In Fig. \ref{fig:apl-fig3}(f), the final position error, max velocity error, and RMS errors were averaged across all combinations of $J$ and $\tau$.  The model performs exceptionally well at most parameter corners.

\begin{figure}
    \centering
    \includegraphics[width=0.48\textwidth]{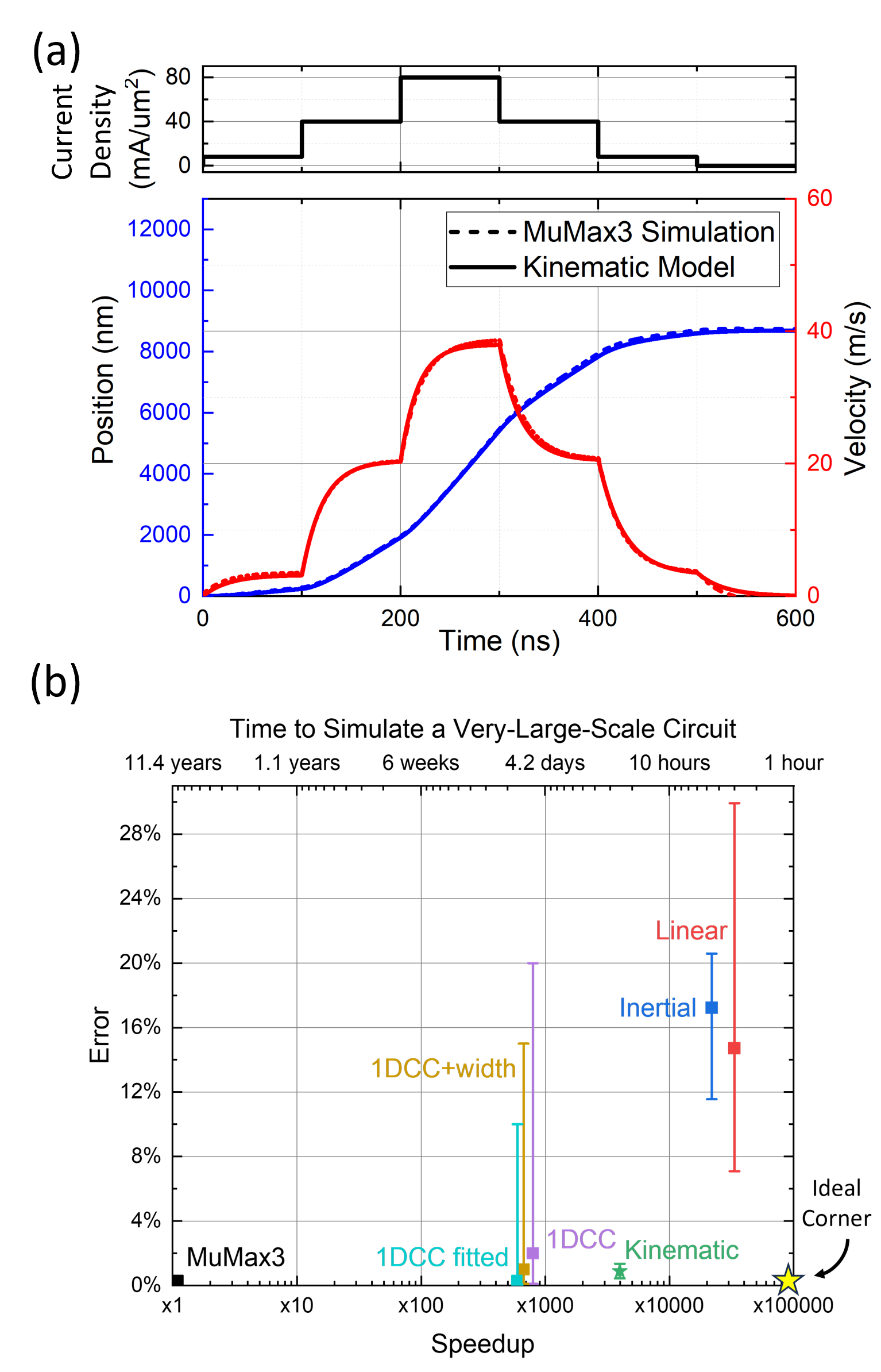}
    \caption{(a) Predictive nature of model for complex input sequences, with Aex = 11x$10^{-12}$, Ku = 4.05x$10^5$, $\alpha$ = 0.01, $M_{sat}$ = 7.95x$10^5$, and track width = 100x$10^{-9}$. (b) Model accuracy and speedup compared relative to previously published work.  Python implementations of the linear model \cite{doi:10.1063/1.3536793}, inertial model \cite{9180675}, kinematic model (this work), 1D CC model with and without variable wall width \cite{9662302}, and 1D CC model with fitting parameters \cite{9072283} are compared.  Speedup is normalized relative to the mumax3 simulations.  Time to simulate a very-large-scale circuit is normalized to one hour.}
    \vspace{-1em}
    \label{fig:apl-fig4}
\end{figure}

Furthermore, the kinematic Verilog-A model demonstrates close matching with more complicated current pulses, indicating the model will be very useful in analog simulations where DW devices are stimulated frequently.  After model fitting, the model and the micromagnetic simulator were presented a complicated piecewise-constant input current signal as illustrated in Fig. \ref{fig:apl-fig4}(a) and Supplementary Movie 1.  The model thus predicts DW position with very high accuracy compared with the micromagnetic simulation.


As illustrated in Fig. \ref{fig:apl-fig4}(b), the proposed kinematic DW model demonstrates significant speedup compared with CC models and significant accuracy boost compared with the oversimplifying models, making it very attractive for analog DW device network simulations. To enable this comparison, the extant DW motion models were implemented in Python (see the Supplementary Information).  For all of these models, both speedup and accuracy were evaluated relative to micromagnetic simulations as a baseline.

Compared with micromagnetic simulations, our proposed model presents a 400x speedup.  This comes at a small accuracy cost of roughly 1.2\% error, on average.  Compared with CC models, our model presents an 7x speedup, with minimal accuracy loss compared with the best CC models.

Finally, our model demonstrates very large accuracy boost compared with the oversimplifying models of \cite{8715734, doi:10.1063/1.3536793}, with an acceptable 50\% loss in speed.  This drastic increase in accuracy is due to the inclusion of non-linear terms which the models of \cite{8715734, doi:10.1063/1.3536793} do not use.



In this work, we present a DW model inspired by the phenomenological similarities between a DW and a classical object subject to Newton's second law of motion.  The model is posted publicly online along with scripts to automatically extract model fitting parameters from new simulation or experimental data.  Over the 32 parameter corners tested in this work, our proposed phenomenological model predicts DW motion within 1.2\% on average compared with micromagnetic simulations.  The model is two orders of magnitude faster than micromagnetic simulations, significantly faster than CC models, and drastically more accurate than overly reductive models.  We therefore anticipate that our proposed kinematic model will significantly advance the simulation and design of large-scale networks of DW devices for digital, analog, and neuromorphic applications.

\section*{References}

\bibliography{main}

\end{document}


\bstctlcite{IEEEexample:BSTcontrol}

\renewcommand{\figurename}{Supplementary Figure}
\renewcommand{\thefigure}{S\arabic{figure}}
\setcounter{figure}{0}

\onecolumn

\begin{center}
{\Large \noindent Supplementary Information for Kinematic Model of Magnetic Domain Wall Motion for Fast, High-Accuracy Simulations}
\end{center}

\normalsize

\vspace{1em}

Kristi Doleh$^{1*}$ Leonard Humphrey$^{1*}$, Chandler M. Linseisen$^{1*}$, Michael D. Kitcher$^2$, Joanna M. Martin$^1$, \\Can Cui$^3$, Jean Anne C. Incorvia$^3$, Felipe Garcia-Sanchez$^{4**}$, Naimul Hassan$^1$, Alexander J. Edwards$^{1**}$, \\Joseph S. Friedman$^{1**}$

$^1$ Electrical and Computer Engineering, The University of Texas at Dallas, Richardson, TX

$^2$ Department of Materials Science \& Engineering, Carnegie Mellon University, Pittsburgh, PA

$^3$ Electrical and Computer Engineering, The University of Texas at Austin, Austin, TX

$^4$ Universidad de Salamanca, Salamanca, Spain

$^*$first authors contributed equally.  $^{**}$e-mail: fgs@usal.es, \{alexander.edwards, joseph.friedman\}@utdallas.edu.

\vspace{1em}

\noindent This document contains:
\begin{itemize}
    \item \hyperref[sec:reduction-argument]{a first-principles justification for the kinematic reduction of DW dynamics,}
    \item \hyperref[sec:micromagnetic-methods]{our micromagnetic simulation methodology,}
    \item \hyperref[sec:fitting-parameters]{methods for choosing fitting parameters,}
    \item \hyperref[sec:speedup]{methods for model speedup and accuracy comparison,}
    \item \hyperref[sec:model]{the Verilog-A kinematic domain wall model,}
    \item \hyperref[sec:table]{the table of micromagnetic and model fitting parameters,}
    \item \hyperref[sec:tutorial]{a tutorial for extending the model to new simulation or experimental data,}
    \item \hyperref[sec:mumax]{the micromagnetic simulation template,}
    \item \hyperref[sec:matlab]{the code to extract model fitting parameters from micromagnetic simulation results, and}
    \item \hyperref[sec:references]{references.}
\end{itemize}
\vspace{1em}
Some lines of code in this document have been broken as signified by ellipsis: `\verb!...!'.
All code presented in this document is publicly available on our Github (\url{https://github.com/AJEdwards314/KinematicDomainWallModel}) and website: (\url{https://personal.utdallas.edu/~joseph.friedman/Papers/KinematicDWModel2023.zip}).   \vspace{5em}

\newpage

\section*{First-Principles Justification for Kinematic Reduction}
\label{sec:reduction-argument}

There is strong evidence that a kinematic DW model is not only appropriate, but also motivated by the first-principles theory.  As can be derived from the Slonczewski equations, the field- or current-driven dynamics of DWs are dictated by forces that closely mirror the classical picture of a harmonic oscillator. The Slonczewski equations below describe the evolution of the DW's position ($x$) and the in-plane angle of its magnetization ($\phi$) in the flow regime where the field is higher than the effective static depinning field. While $\phi$ has no classical analog, it can be considered as a hidden degree of freedom that modulates the relevant forces as well as the mass of the DW.

Limiting ourselves to the 1-dimensional CC form of the Slonczewski equations \cite{Malozemoff1979, Emori2013}, we have
\begin{subequations}
\begin{align}
    \dot{\phi} &= {\frac{\gamma}{1 + \alpha^2}}\left(-\alpha\Omega - B_c{sgn[\dot{x}]} + \frac{\pi}{2}B_{SH}\cos[\phi]\right)\label{eq:phiDot};\\
    \dot{x} &= {\frac{\gamma\lambda_0}{1 + \alpha^2}}\left(\Omega - {\alpha}B_c{sgn[\dot{x}]} + \alpha\frac{\pi}{2}{B_{SH}}\cos[\phi]\right)\label{eq:qDot};\\
    \Omega &= \frac{\sigma_{\phi}}{{2\lambda_0}M_s};\\
    B_{SH} &= \frac{\hbar\theta_{SH}J}{2eM_s{t_f}}\label{eq:restoringTorque}.
\end{align}
\end{subequations}
Here $\gamma$ is the gyromagnetic ratio, $\lambda_0$ is the DW width, $\alpha$ is the Gilbert damping, $M_s$ is the saturation magnetization, $\sigma$ is the energy per cross-sectional area of the DW, $B_c$ is a static DW coercive field (we ignore other pinning mechanisms), $\hbar$ is the reduced Planck's constant,  $\theta_{SH}$ is the effective spin Hall angle, $e$ is the elementary charge, $J$ is the current density and $B_{SH}$ is the current-induced effective field from the Slonczewski-like torque due to the spin Hall effect.

To obtain the transient dynamics of the DW, \eqref{eq:phiDot} and \eqref{eq:qDot} can be linearized about the equilibrium magnetization angle $\phi_{eq}$ \cite{Malozemoff1979, Torrejon2016}. Assuming small driving currents and, therefore, small deviations from the static DW orientation  (\textit{i.e.}, $\sin[\phi] \approx \phi$ and $\phi^2 \approx 0$) results in the following harmonic oscillator equation:

\begin{subequations}
\begin{align}
   \frac{F}{m} &= \frac{{\partial^2}x}{\partial{t^2}} + \frac{1}{\tau}\frac{\partial{x}}{\partial{t}}\label{eq:SHO};\\ 
   F &= 2\pi{w{t_f}}M_s\left({B_{SH}}{\cos[\phi_{eq}]}-B_c{sgn[\dot{x}]}\right)\label{eq:Force};\\
   m &= \left(\frac{2M_s}{\gamma}\right)^2\frac{(1 +\alpha^2)w{t_f}}{f\left[\phi_{eq}\right]}\nonumber\\
   &= \frac{m_0}{f\left[\phi_{eq}\right]}\label{eq:mass};\\
   \tau &= 
   \frac{\lambda_0}{\alpha}\left(\frac{2M_s}{\gamma}\right)\frac{(1 +\alpha^2)}{f\left[\phi_{eq}\right]}\label{eq:tau};\\
   f\left[\phi_{eq}\right] &= {\sigma_{\phi\phi}|}_{\phi_{eq}} - \frac{{\lambda_0}}{\alpha}\pi{M_s}{B_{SH}}{\sin[\phi_{eq}]}.
\end{align}
\end{subequations}
Here, $F$ is the current-induced driving force, $\tau$ is the relaxation time of the transient dynamics, $m$ is the effective mass of the DW, $w$ and $t_f$ are the DW length and film thickness, respectively, $m_0$ is the $\phi$-independent component of the DW mass.  Rearranging \eqref{eq:SHO} and grouping terms, we get:

\begin{equation}
    \frac{\partial^2x}{\partial t^2} = f_1(J)\frac{\partial x}{\partial t} + f_2(J), 
    \label{eq.3}
\end{equation}
which is identical in form to the motion of an object under Stokes' drag as illustrated in Fig. 1 of the paper; DW motion can therefore be efficiently and accurately modeled as an object under drag.

$f_1$ and $f_2$ have a complex dependence on $J$, as $\phi[t]$ also changes with $J$ according to \eqref{eq:phiDot}. From \eqref{eq:SHO}, this driving force is balanced by an inertial force proportional to $m$ and a viscous-damping term proportional to the velocity. Finally, DW motion only occurs when the current-induced force is higher than the static pinning force, $F_c \propto B_c$. Although this framework can be augmented to account for additional phenomena (\textit{e.g.} DW tilting \cite{Boulle2013}, dependence of DW width on current \cite{Gehanne2021}, dynamical depinning \cite{Moretti2017}, large deviations from the static $\phi$ \cite{Torrejon2016}), the prevailing forces at play in the flow regime remain the same, indicating that they can be simply modeled using second-order kinematic differential equations.

\section*{Micromagnetic Simulation Methodology}
\label{sec:micromagnetic-methods}
Track width is varied between 50 nm and 100 nm. Track length is 500 nm; as the $ext\_centerWall$ command is used, the simulation window shifts to keep the DW in the center of the simulation to emulate an infinitely long track.  Simulation cells are 1 nm by 1.6 nm by 1.2 nm, and track thickness is 1.2 nm.
Simulations were run across a wide range of parameter corners for track width, $M_{sat}, B_{anis,eff}, \alpha,$ and $Aex$, with $Ku$ determined from $M_{sat}$ and $B_{anis,eff}$ according to $Ku = \frac{1}{2} M_{sat} (\mu_0 M_{sat} + B_{anis,eff}) \hat{k}$.  For each parameter corner, current densities were specifically chosen to avoid Walker breakdown. The interfacial Dzyaloshinskii-Moriya strength (Dind) was set to $0.25$  mJ/m$^2$. A Bloch DW is instantiated at the center of the simulation.  SOT was emulated in mumax3 as a small z-directed STT current with a spin-Hall proportion of 0.05.

The mumax3 $ext\_centerWall$ command is used to simulate a track of infinite length, and the DW position is recorded.  However, as mumax3 reports the DW position ($ext\_dwpos$) as an integer, a slight offset is used to smooth the DW position trace.  This offset is determined by finding the maximum magnetization change in the track.  DW velocity is extracted as the first order discrete difference of DW position over time.

\section*{Choosing Fitting Parameters}
\label{sec:fitting-parameters}

In order to choose the seven fitting parameters of our model for each parameter corner, the following quantities were extracted from micromagnetic simulations and analyzed: maximum velocity, the drift distance after current was turned off, and the time constant of the velocity increase given a constant current.  

The relationship between maximum velocity and current density was fit to a cubic polynomial, where $v_{max} = c_0 + c_1J + c_2J^2 + c_3J^3$. The Matlab function $fitnlm$ was used to fit to this polynomial, with a weight ensuring closer fit to lower values of max velocity.

The relationship between the drift distance (with no applied current) and the corresponding initial velocity (max velocity) was fit to a linear model, where $driftDist = driftConst * v_{max}$. The function $fitnlm$ from Matlab was used to perform this fit, with a weight ensuring a close fit to lower values of drift distance. The model parameter corresponding to this fit is $d_1$, where $d_1 = 1/driftConst$.

The relationship between the velocity time constant -- defined as the time interval from the start of a current pulse to the point where the velocity reaches $63.2\%$ ($1-1/e$) of its maximum value -- and the current density was fit to a rational relationship, where $\tau = \frac{1}{d_2J + d_1}$. This fit was performed by finding the least squares fit between the current and the inverse time constant.

The equations used for fitting were found by solving our model equations for these three relationships. As a result, the resulting best fit parameters have an exact correspondence to the model parameters, described by the following equations:

\begin{equation*}
\begin{array}{l}
    k_0 = d_1*c_0,\\

    k_1 = d_1*c_1 + d_2*c_0, \\

    k_2 = d_1*c_2 + d_2*c_1, \\

    k_3 = d_1*c_3 + d_2*c_2, \text{and} \\

   k_4 = d_2*c_3.
\end{array}
\end{equation*}

\section*{Model Speedup and Accuracy Comparison}
\label{sec:speedup}

To compare the speedup and accuracy of the proposed kinematic model, we implemented the models of \cite{doi:10.1063/1.3536793, 9180675, 9662302, 9072283} and the proposed model in Python.  All models were simulated with a discrete time-step of 1 ps, for a duration of 6 $\mu$s of simulation time.  All Python models were run on a machine with an Intel i9-9920X processor, and speedup and accuracy were compared against Mumax3 simulations that performed were with an NVIDIA GeForce RTX 2080 GPU Ti.  The authors had difficulty getting the CC models of \cite{9662302, 9072283} to converge, so the relative CC accuracies presented in Fig. 4b of the paper are best-guess estimates, with the point of interest being the speedup of the proposed model over the CC models.  Model algorithms are outlined as follows in order of increasing complexity:

\begin{breakablealgorithm}
\caption{Linear Model \cite{doi:10.1063/1.3536793}}
\label{alg:cap}
\begin{algorithmic}

\State define time array
\State define current density array as $J$
\State $\beta \gets$ value from lookup table
\State $x \gets$ 0 array
\State $v \gets$ 0 array

\State start timer
\For{each time $t$ in time array}
    \State $v$ appends $\beta * J(t)$ 
    \State $x$ appends $x_{prev} + v_{prev} * \Delta t$
\EndFor
\State end timer

\end{algorithmic}
\end{breakablealgorithm}

\begin{breakablealgorithm}
\caption{Inertial Linear Model \cite{9180675}}
\begin{algorithmic}

\State define time array
\State define current density array as $J$
\State $\beta \gets$ value from lookup table
\State $X_{dw} \gets W_{DW}/2\alpha$
\State $X_{dw,cache} \gets 0$ 
\State $x \gets$ 0 array
\State $v \gets$ 0 array

\State start timer
\For{each time $t$ in time array}
    \If{$J(t) > 2x10^8$}    
        \State $v$ appends $-\beta * J(t)$ 
        \State $X_{dw,cache} \gets X_{dw}$
    \ElsIf{$X_{dw,cache}>0$}
        \State $X_{dw,cache} \gets X_{dw,cache}-|v_{prev}|* \Delta t$
        \State $v$ appends $v_{prev}$
    \Else
        \State $v$ appends $0$
    \EndIf
     \State $x$ appends $x_{prev} + v_{prev} * \Delta t$

\EndFor

\State end timer

\end{algorithmic}
\end{breakablealgorithm}

\begin{algorithm}
\caption{Kinematic Model}
\begin{algorithmic}

\State define time array
\State define current density array as $J$
\State $c_0, c_1, c_2, c_3, driftConst, d_2 \gets $ input from lookup table

\State $d_1 = 1/driftConst$

\State $k_0 \gets d_1*c_0$
\State $k_1 \gets d_1*c_1 + d_2*c_0$
\State $k_2 \gets d_1*c_2 + d_2*c_1$
\State $k_3 \gets d_1*c_3 + d_2*c_2$
\State $k_4 \gets d_2*c_3$

\State $x \gets$ 0 array
\State $v \gets$ 0 array
\State $a \gets$ 0 array

\State start timer
\For{each time $t$ in time array}
    \If{$J(t) > 2e8$}    
        \State $J_{sign} \gets sign[J(t)]$  
    \Else
        \State $J_{sign} \gets 0$  
    \EndIf
    \State $a$ appends $-J_{sign} * [k_0 + k_4 |J(t)|^4 + k_3 |J(t)|^3 + k_2 |J(t)|^2 + k_1 |J(t)|] - [d_1 + d_2 |J(t)|] * v_{prev}$
    \State $v$ appends $v_{prev} + a_{prev} * \Delta t$  [in ns]
     \State $x$ appends $x_{prev} + v_{prev} * \Delta t$ [in s]
\EndFor
\State end timer

\end{algorithmic}
\end{algorithm}

\begin{breakablealgorithm}
\caption{Collective coordinate model (with constant wall width) \cite{9662302}}
\begin{algorithmic}

\State define time array
\State define current density array as $J$

\State $W_{DW} \gets \sqrt{A_{ex}/(K_u-\mu_0 M_{sat}^2/2)}$
\State $N_x \gets t_{track}/(t_{track}+W_{DW})$
\State $N_y \gets t_{track}/(t_{track}+W_{track})$
\State $H_k \gets (N_x - N_y)M_{sat}$
\State $H_x, H_y, H_z \gets 0$
\State $H_D \gets Dind/(\mu_0 M_{sat} W_{DW})$

\State $q \gets$ 0 array
\State $\phi \gets$ 0 array
\State $\chi \gets$ 0 array

\State start timer
\For{each time $t$ in time array}
    \State $H_{SHE} \gets \mu_B \theta_{SH} J(t)/(\gamma_0 e M_{sat} t_{track})$
    \State $\sigma \gets (1 /\cos(\chi_{prev})) (4\sqrt{\alpha K_u} - Q\pi D_{ind}\cos(\phi_{prev} - \chi_{prev}) + \mu_0 H_K M_{sat} W_{DW} \cos^2[2(\phi_{prev} - \chi_{prev})])$
    \State $\Omega_A \gets -\frac{1}{2} \gamma_0 H_K \sin[2(\phi_{prev} - \chi_{prev})] - \frac{\pi}{2} \gamma_0 H_y * \cos(\phi_{prev}) + \frac{\pi}{2} \gamma_0 H_x \sin(\phi_{prev}) + \frac{\pi}{2} \gamma_0 H_D \sin(\phi_{prev} - \chi_{prev})$
    \State $\Omega_B \gets \gamma_0 Q H_z + \frac{\pi}{2} \gamma_0 Q H_{SHE} \cos(\phi_{prev})$
    \State $\Omega_C \gets -\sigma \sin(\chi_{prev}) + \pi D_{ind} Q \sin(\phi_{prev} - \chi_{prev}) - \mu_0 H_K M_{sat} W_{DW} \sin[2(\phi_{prev} - \chi_{prev})]$
    
    \State $dq \gets \Delta t * \frac{W_{DW} (\Omega_A + \alpha \Omega_B)}{ \cos(\chi_{prev})*(1+\alpha^2)}$
    \State $d\phi \gets \Delta t * \frac{\Omega_B-\alpha\Omega_A}{1+\alpha^2}$
    \State $d\chi \gets \Delta t *\frac{6 \gamma_0 \Omega_C/(\alpha \mu_0 M_{sat} W_{DW} \pi^2)}{ \tan^2(\chi_{prev}) + (W_{track}/\pi W_{DW} \cos\chi_{prev})^2} $
    \State $q$ appends $q_{prev} + dq$
    \State $\phi$ appends $\phi_{prev} + d\phi$
    \State $\chi$ appends $\chi_{prev} + d\chi$
    
\EndFor
\State end timer

\end{algorithmic}
\end{breakablealgorithm}

\begin{breakablealgorithm}
\caption{Collective coordinate model (with variable wall width) \cite{9662302}}
\begin{algorithmic}

\State define time array
\State define current density array as $J$

\State $W_{DW} \gets \sqrt{A_{ex}/(K_u-\mu_0 M_{sat}^2/2)}$
\State $N_x \gets t_{track}/(t_{track}+W_{DW})$
\State $N_y \gets t_{track}/(t_{track}+W_{track})$
\State $N_z \gets 1 - N_x - N_y$
\State $H_k \gets (N_x - N_y)M_{sat}$
\State $H_x, H_y, H_z \gets 0$
\State $H_D \gets Dind/(\mu_0 M_{sat} W_{DW})$

\State $q \gets$ 0 array
\State $\phi \gets$ 0 array
\State $\chi \gets$ 0 array

\State start timer
\For{each time $t$ in time array}
    \State $H_{SHE} \gets \mu_B \theta_{SH} J(t)/(\gamma_0 e M_{sat} t_{track})$
    \State $\sigma \gets (1 /\cos(\chi_{prev})) (4\sqrt{\alpha K_u} - Q\pi D_{ind}\cos(\phi_{prev} - \chi_{prev}) + \mu_0 H_K M_{sat} W_{DW} \cos^2[2(\phi_{prev} - \chi_{prev})])$
    \State $\Omega_A \gets -\frac{1}{2} \gamma_0 H_K \sin[2(\phi_{prev} - \chi_{prev})] - \frac{\pi}{2} \gamma_0 H_y * \cos(\phi_{prev}) + \frac{\pi}{2} \gamma_0 H_x \sin(\phi_{prev}) + \frac{\pi}{2} \gamma_0 H_D \sin(\phi_{prev} - \chi_{prev})$
    \State $\Omega_B \gets \gamma_0 Q H_z + \frac{\pi}{2} \gamma_0 Q H_{SHE} \cos(\phi_{prev})$
    \State $\Omega_C \gets -\sigma \sin(\chi_{prev}) + \pi D_{ind} Q \sin(\phi_{prev} - \chi_{prev}) - \mu_0 H_K M_{sat} W_{DW} \sin[2(\phi_{prev} - \chi_{prev})]$
    
    \State $dq \gets \Delta t * \frac{W_{DW} (\Omega_A + \alpha \Omega_B)}{ \cos(\chi_{prev})*(1+\alpha^2)}$
    \State $d\phi \gets \Delta t * \frac{\Omega_B-\alpha\Omega_A}{1+\alpha^2}$
    \State $d\chi \gets \Delta t *\frac{6 \gamma_0 \Omega_C/(\alpha \mu_0 M_{sat} W_{DW} \pi^2)}{ \tan^2(\chi_{prev}) + (W_{track}/\pi W_{DW} \cos\chi_{prev})^2} $
    \State $q$ appends $q_{prev} + dq$
    \State $\phi$ appends $\phi_{prev} + d\phi$
    \State $\chi$ appends $\chi_{prev} + d\chi$
    
    \State $\delta \gets N_x \cos^2(\phi_{prev}-\chi_{prev}) + N_y\sin^2(\phi_{prev}-\chi_{prev})-N_z$
    \State $W_{DW} \gets \sqrt{A_{ex} / (K_u + \delta \mu_0 M_{sat}^2 / 2)}$
    \State $N_x \gets t_{track}/(t_{track}+W_{DW})$
    \State $N_z \gets 1 - N_x - N_y$
    \State $H_k \gets (N_x - N_y)M_{sat}$
    \State $H_D \gets Dind/(\mu_0 M_{sat} W_{DW})$
    
\EndFor
\State end timer

\end{algorithmic}
\end{breakablealgorithm}

\begin{breakablealgorithm}
\caption{Collective coordinate model with fitting parameters \cite{9072283}}
\begin{algorithmic}

\State define time array
\State define current density array as $J$

\State $W_{DW} \gets \sqrt{A_{ex}/(K_u-\mu_0 M_{sat}^2/2)}$
\State $N_x \gets t_{track}/(t_{track}+W_{DW})$
\State $N_y \gets t_{track}/(t_{track}+W_{track})$
\State $N_z \gets 1 - N_x - N_y$
\State $H_k \gets (N_x - N_y)M_{sat}$
\State $H_x, H_y, H_z \gets 0$
\State $H_D \gets Dind/(\mu_0 M_{sat} W_{DW})$

\State $q \gets$ 0 array
\State $\phi \gets$ 0 array
\State $\chi \gets$ 0 array

\State start timer
\For{each time $t$ in time array}
    \State $H_{SHE} \gets \mu_B \theta_{SH} J(t)/(\gamma_0 e M_{sat} t_{track})$
    \State $I_1 \gets 0.98 + 2/[(W_{track}/W_{DW})^2+2]$
    \State $I_2 \gets 1 +  \frac{W_{track}/W_{DW}}{15(W_{track}/W_{DW}) + 100}$
    \State $\sigma \gets (1 /\cos(\chi_{prev})) (4\sqrt{\alpha K_u} - I_1 Q\pi D_{ind}\cos(\phi_{prev} - \chi_{prev}) + \mu_0 H_K M_{sat} W_{DW} \cos^2[2(\phi_{prev} - \chi_{prev})])$
    \State $\Omega_A \gets -\frac{1}{2} \gamma_0 H_K \sin[2(\phi_{prev} - \chi_{prev})] - \frac{\pi}{2} \gamma_0 H_y * \cos(\phi_{prev}) + \frac{\pi}{2} \gamma_0 H_x \sin(\phi_{prev}) + I_1\frac{\pi}{2} \gamma_0 H_D \sin(\phi_{prev} - \chi_{prev})$
    \State $\Omega_B \gets \gamma_0 Q H_z + I_2\frac{\pi}{2} \gamma_0 Q H_{SHE} \cos(\phi_{prev})$
    \State $\Omega_C \gets -\sigma \sin(\chi_{prev}) + \pi D_{ind} Q \sin(\phi_{prev} - \chi_{prev}) - \mu_0 H_K M_{sat} W_{DW} \sin[2(\phi_{prev} - \chi_{prev})]$
    
    \State $dq \gets \Delta t * \frac{W_{DW} (\Omega_A + \alpha \Omega_B)}{ \cos(\chi_{prev})*(1+\alpha^2)}$
    \State $d\phi \gets \Delta t * \frac{\Omega_B-\alpha\Omega_A}{1+\alpha^2}$
    \State $d\chi \gets \Delta t *\frac{6 \gamma_0 \Omega_C/(\alpha \mu_0 M_{sat} W_{DW} \pi^2)}{ \tan^2(\chi_{prev}) + (W_{track}/\pi W_{DW} \cos\chi_{prev})^2} $
    \State $q$ appends $q_{prev} + dq$
    \State $\phi$ appends $\phi_{prev} + d\phi$
    \State $\chi$ appends $\chi_{prev} + d\chi$
    
    \State $\delta \gets N_x \cos^2(\phi_{prev}-\chi_{prev}) + N_y\sin^2(\phi_{prev}-\chi_{prev})-N_z$
    \State $W_{DW} \gets \sqrt{A_{ex} / (K_u + \delta \mu_0 M_{sat}^2 / 2)}$
    \State $N_x \gets t_{track}/(t_{track}+W_{DW})$
    \State $N_z \gets 1 - N_x - N_y$
    \State $H_k \gets (N_x - N_y)M_{sat}$
    \State $H_D \gets Dind/(\mu_0 M_{sat} W_{DW})$
    
\EndFor
\State end timer

\end{algorithmic}
\end{breakablealgorithm}

\newpage

\section*{VerilogA Model}
\label{sec:model}

\begin{verbatim}
    
`include "constants.vams"
`include "disciplines.vams"

nature distance 
  access = Metr;
  units = "m";
  abstol = 0.01n;
endnature
 
discipline Distance
  potential distance;
enddiscipline

module DW_kinematic_model(P, Q, DW_Position, RA, DW_Velocity, DW_MC, DW_Req,DW_J,...
    A_J,A_d1,A_d2);

inout P, Q;					//P is defined as the left port of the DW track, and Q is the right port.
inout RA;					//MTJ output port
output DW_Position;				//This port measures the DW position, in meters
output DW_Velocity;				//This port measures the DW velocity, in meters
output DW_MC;					//This port measures the effective DW track current, in Amps.
output DW_J;					//This port measures the effective DW track current density, in Amps.

// Ports measuring various acceleration contributions, for tracking
output A_J;
output A_d1;
output A_d2;

electrical P, Q;
electrical RA;
electrical Vmiddle;

Distance DW_Position;
Distance DW_Velocity;
Distance DW_MC;
Distance DW_J;		
Distance DW_Req;

parameter real I_th = 1n;			// threshold current for DW motion				

// Resistance parameters
parameter real Rp = 1k;
parameter real Rap = 1M;
parameter real R_init = 10k;
parameter real Rtotal = 3.9k;

// DW track length and MTJ position
parameter real Pdw_init = 0;
parameter real Pdw_limit = 120n;
parameter real Pdw_low = 20n;
parameter real Pdw_high = 40n;

//------------------------------------MTJ resistance voltage dependence section

parameter real Switch_MTJ_VoltageDependent = 0;
parameter real MTJ_VoltageDependent_Factor = 1;
parameter real MTJ_VoltageDependent_MinValue = 0.4; 

//------------------------------------MTJ resistance voltage dependence section end

parameter real dt = 1e-3;			// time step (in ns)

// micromagnetic parameters
parameter real Aex = 11e-12;
parameter real Ku = 9.17e5;
parameter real Msat = 1.2e6;
parameter real alpha = 0.01;
parameter real DW_width = 50e-9;
parameter real DW_thickness = 1.2e-9;
parameter real Areaa = DW_thickness*DW_width;		// cross section area of heavy metal layer
parameter real thetaSH = 0.05;
real B_anis = 0;				// B_anis calculated from other params

parameter real c_r = 0.25;			// coefficient of restitution, controls track end ...
    bounce, between 0 and 1

// Domain Wall motion variables
real Pdw;				//measured in m, not nm
real Vdw;
real Adw;

// Resistance variables
real Req;
real RR;
real RL;

// Current density variables
real Jdenss;
real J_sign;
real J_init;
real IPQPQ;
real prev_time;

// model constants
real c0; //Change to "parameter real" to set manually.
real c1; //Change to "parameter real" to set manually.
real c2; //Change to "parameter real" to set manually.
real c3; //Change to "parameter real" to set manually.
real drift_const; //Change to "parameter real" to set manually.

real k0;
real k1;
real k2;
real k3;
real k4;
real d1;
real d2; //Change to "parameter real" to set manually.

Distance A_J;
Distance A_d1;
Distance A_d2;

//------------------------------------MTJ resistance voltage dependence section

real Rp_eq;
real Rap_eq;
real VMTJ_eq; 

//------------------------------------MTJ resistance voltage dependence section end

analog
begin
	@(initial_step)
	begin

		Rp_eq = Rp;
		Rap_eq = Rap;


		Pdw = Pdw_init;
		Vdw = 0;

		Jdenss = 0;
		J_init = 0;
		
		
		//Req = (((Pdw_limit - Pdw) / Pdw_limit ) * Rap) + ((Pdw / Pdw_limit) * Rp) ; 		
		Req = R_init;

		if(Ku == 1.11e6 || Ku == 5.36e5)
		begin
    			B_anis = 350;
		end
		else if(Ku == 9.17e5 || Ku == 4.05e5)
		begin
    			B_anis = 20;
		end
		else if(Ku == 7.01e5)
		begin
			B_anis = 150;
		end
		else
		begin
    			B_anis = (Ku/(0.5*Msat) - (4*`M_PI*1e-7)*Msat)*1000;
		end

//Comment out the following block if constants c0-c3, drift_const, and d2 manually set.
		//Read in constant values for specific parameter corner from lookup table files ...
    (included on github).
		c0 = $table_model (Aex*1e12, B_anis,alpha,Msat,DW_width*1e9, ...
    "lookup_maxVel_c0.tbl", "1L,1L,1L,1L,1L");
		c1 = $table_model (Aex*1e12, B_anis,alpha,Msat,DW_width*1e9, ...
    "lookup_maxVel_c1.tbl", "1L,1L,1L,1L,1L");
		c2 = $table_model (Aex*1e12, B_anis,alpha,Msat,DW_width*1e9, ...
    "lookup_maxVel_c2.tbl", "1L,1L,1L,1L,1L");
		c3 = $table_model (Aex*1e12, B_anis,alpha,Msat,DW_width*1e9, ...
    "lookup_maxVel_c3.tbl", "1L,1L,1L,1L,1L");
		drift_const = $table_model (Aex*1e12, B_anis,alpha,Msat,DW_width*1e9, ...
    "lookup_drift_const.tbl", "1L,1L,1L,1L,1L");
		d2 = $table_model (Aex*1e12, B_anis,alpha,Msat,DW_width*1e9, ...
    "lookup_d2.tbl", "1L,1L,1L,1L,1L");
		
		// calculate other constants
		d1 =  1/drift_const; 
		k0 = d1*c0;
    		k1 = d1*c1 + d2*c0;
    		k2 = d1*c2 + d2*c1;
    		k3 = d1*c3 + d2*c2;
    		k4 = d2*c3;

	end

	//if pos of DW below (left of) MTJ, current through wall determined by left pin ...
    (P) and MTJ voltage
	if(Pdw < ((Pdw_low + Pdw_high)/2))
	begin
		IPQPQ = I(P, Vmiddle);
	end
	else
	begin
		IPQPQ = I(Vmiddle, Q);
	end
	
	//Conversion of current through heavy metal layer to current density
	if(abs(IPQPQ) >= I_th)
	begin
		Jdenss = thetaSH*(IPQPQ / Areaa);
	end
	else
	begin
		Jdenss = 0;
	end
                          
	// find sign of current density
	J_sign = 0;
	if(Jdenss >0)
	begin
		J_sign = 1;
	end
	else if(Jdenss<0)
	begin
		J_sign = -1;
	end
	else
	begin
		J_sign = 0;
	end

	// Calculation of acceleration, velocity, position
	Adw = - J_sign*(k0 + k4*(abs(Jdenss)**4) + k3*(abs(Jdenss)**3) + k2* ...
    (abs(Jdenss)**2) + k1*abs(Jdenss)) - (d1 + d2*abs(Jdenss))*Vdw;	// [nm/ns^2]
	Vdw = Vdw + Adw * dt;															// [nm/ns = m/s]
	Pdw = Pdw + Vdw * dt*1e-9;														// [m]
	
//---------------------------------------Edge bounce

	if(Pdw<=0)
	begin
		Pdw = 0;
		Vdw = -c_r*Vdw;
	end
	else if(Pdw>=Pdw_limit)
	begin
		Pdw = Pdw_limit;
		Vdw = -c_r*Vdw;
	end

//------------------------------------MTJ resistance voltage dependence section

	if(Switch_MTJ_VoltageDependent == 1)
	begin
	
		VMTJ_eq = V(RA, Vmiddle);

		Rap_eq = (1 - VMTJ_eq * MTJ_VoltageDependent_Factor) * Rap;

		if(Rap_eq < MTJ_VoltageDependent_MinValue * Rap)
		begin
			Rap_eq = MTJ_VoltageDependent_MinValue * Rap;
		end

	end

//------------------------------------MTJ resistance voltage dependence section end

	if(Pdw <= Pdw_low && Pdw >=0)
	begin
		Req = Rp_eq;
	end
	else if(Pdw >= Pdw_high && Pdw <= Pdw_limit)
	begin
		Req = Rap_eq;
	end
	else
	begin 

		Req = ((Pdw_high - Pdw_low) * Rp_eq * Rap_eq) / ((Rp_eq * (Pdw - Pdw_low)) + ...
    ( Rap_eq * (Pdw_high - Pdw)));
	end

	RR = (Pdw_limit-Pdw_low-10n)/(Pdw_limit)*Rtotal;
	RL = Rtotal - RR;

	// calculate currents through device
	I(P, Vmiddle) <+ V(P, Vmiddle) / RL;
	I(Q, Vmiddle) <+ V(Q, Vmiddle) / RR;

	I(Vmiddle, RA) <+ V(Vmiddle, RA) / Req;

	// Components of acceleration split up, measurable on ports
	Metr(A_J) <+ - J_sign*(k0 + k4*(abs(Jdenss)**4) + k3*
    (abs(Jdenss)**3) + k2*(abs(Jdenss)**2) + k1*abs(Jdenss));
	Metr(A_d1) <+ -d1*Vdw;
	Metr(A_d2) <+ - d2*abs(Jdenss)*Vdw;

	// domain wall motion variables, current and current density measurable on ports
	Metr(DW_Req) <+ Req;
	Metr(DW_Position) <+ Pdw;
	Metr(DW_Velocity) <+ Vdw;
	Metr(DW_MC) <+ IPQPQ;
	Metr(DW_J) <+ Jdenss/1e9;

end
endmodule

\end{verbatim}

\newpage

\section*{Table of Fitting Parameters}
\label{sec:table}

\begin{table}[b!]
    \begin{tabular}{ccccc|cccccc}
\textbf{Aex}&\textbf{B}$_{\mathbf{anis}}$&$\mathbf{\alpha}$&\textbf{Msat}&\textbf{W}&&&&&\textbf{drift}&\\
pJ/m&mT&&kA/m&nm&\textbf{c$_0$}&\textbf{c$_1$}&\textbf{c$_2$}&\textbf{c$_3$}&\textbf{const}&\textbf{d$_2$}\\\hline

\rowcolor{gray!25} 11 & 350 & 0.01 & 1200 & 100 & -0.0406631 & 1.94592x$10^{-09}$ & -4.19827x$10^{-20}$ & 3.83265x$10^{-31 }$ & 19.8039 & 1.29749x$10^{-12}$  \\ 
\rowcolor{gray!25} 11 & 350 & 0.01 & 1200 & 50  & -0.749487  & 2.24379x$10^{-09}$ & -4.11644x$10^{-20}$ & 2.92975x$10^{-31 }$ & 13.8068 & 1.79247x$10^{-12}$  \\ 
\rowcolor{white}   11 & 350 & 0.05 & 1200 & 100 & -0.109504  & 4.00796x$10^{-10}$ & -1.04641x$10^{-21}$ & -1.55942x$10^{-32}$ & 5.98221 & 1.14876x$10^{-13}$  \\ 
\rowcolor{white}   11 & 350 & 0.05 & 1200 & 50  & -0.0366871 & 4.02894x$10^{-10}$ & -2.67686x$10^{-22}$ & -1.24944x$10^{-32}$ & 2.91603 & 3.54164x$10^{-14}$  \\ 
\rowcolor{gray!25} 11 & 20  & 0.01 & 795  & 100 & -1.58242   & 1.2232x$10^{-08 }$ & -6.24891x$10^{-19}$ & 1.04196x$10^{-29 }$ & 26.9651 & 5.00925x$10^{-12}$  \\ 
\rowcolor{gray!25} 11 & 20  & 0.01 & 795  & 50  & -1.45563   & 1.08742x$10^{-08}$ & -3.5832x$10^{-19 }$ & 3.05929x$10^{-30 }$ & 14.9599 & 4.91715x$10^{-12}$  \\ 
\rowcolor{white}   11 & 20  & 0.05 & 795  & 100 & -0.597257  & 2.34428x$10^{-09}$ & -1.78466x$10^{-20}$ & -6.29929x$10^{-32}$ & 5.11551 & 2.39672x$10^{-12}$  \\ 
\rowcolor{white}   11 & 20  & 0.05 & 795  & 50  & -0.146572  & 1.94896x$10^{-09}$ & -3.06343x$10^{-21}$ & -1.28947x$10^{-31}$ & 2.77719 & 2.05405x$10^{-12}$  \\ 
\rowcolor{gray!25} 11 & 350 & 0.01 & 795  & 100 & -0.895302  & 4.39822x$10^{-09}$ & -1.09404x$10^{-19}$ & 1.03521x$10^{-30 }$ & 13.2439 & 3.04999x$10^{-12}$  \\ 
\rowcolor{gray!25} 11 & 350 & 0.01 & 795  & 50  & -1.76883   & 4.602x$10^{-09  }$ & -9.96213x$10^{-20}$ & 7.85948x$10^{-31 }$ & 10.5141 & 3.51861x$10^{-12}$  \\ 
\rowcolor{white}   11 & 350 & 0.05 & 795  & 100 & -0.0819332 & 7.98112x$10^{-10}$ & -1.21205x$10^{-21}$ & -3.19612x$10^{-32}$ & 2.69524 & 1.11848x$10^{-12}$  \\ 
\rowcolor{white}   11 & 350 & 0.05 & 795  & 50  & -0.0342627 & 7.78804x$10^{-10}$ & -3.36143x$10^{-22}$ & -2.63756x$10^{-32}$ & 1.93966 & 1.55811x$10^{-12}$  \\ 
\rowcolor{gray!25} 11 & 20  & 0.01 & 1200 & 100 & 1.01578    & 4.7374x$10^{-09 }$ & -1.60047x$10^{-19}$ & 1.79483x$10^{-30 }$ & 30.8337 & 3.29422x$10^{-12}$  \\ 
\rowcolor{gray!25} 11 & 20  & 0.01 & 1200 & 50  & -0.753876  & 4.71101x$10^{-09}$ & -9.49549x$10^{-20}$ & 3.97306x$10^{-31 }$ & 14.0762 & 3.38823x$10^{-12}$  \\ 
\rowcolor{white}   11 & 20  & 0.05 & 1200 & 100 & -0.248576  & 1.01577x$10^{-09}$ & -4.21517x$10^{-21}$ & -4.83106x$10^{-32}$ & 5.41699 & 9.63854x$10^{-13}$  \\ 
\rowcolor{white}   11 & 20  & 0.05 & 1200 & 50  & -0.0742695 & 8.6009x$10^{-10 }$ & -2.79387x$10^{-22}$ & -3.4938x$10^{-32 }$ & 2.5394  & 3.90235x$10^{-13}$  \\ 
\rowcolor{gray!25} 31 & 350 & 0.01 & 1200 & 100 & -0.582192  & 3.77846x$10^{-09}$ & -1.04774x$10^{-19}$ & 1.04757x$10^{-30 }$ & 23.9387 & 2.1304x$10^{-12 }$  \\ 
\rowcolor{gray!25} 31 & 350 & 0.01 & 1200 & 50  & -2.03832   & 4.15387x$10^{-09}$ & -7.86028x$10^{-20}$ & 4.20631x$10^{-31 }$ & 13.8803 & 2.81582x$10^{-12}$  \\ 
\rowcolor{white}   31 & 350 & 0.05 & 1200 & 100 & -0.150961  & 7.29372x$10^{-10}$ & -2.02797x$10^{-21}$ & -2.85414x$10^{-32}$ & 5.06097 & 3.7673x$10^{-13 }$  \\ 
\rowcolor{white}   31 & 350 & 0.05 & 1200 & 50  & -0.0560668 & 6.96839x$10^{-10}$ & -1.0105x$10^{-22 }$ & -2.00951x$10^{-32}$ & 2.40106 & 1.11262x$10^{-13}$  \\ 
\rowcolor{gray!25} 31 & 20  & 0.01 & 795  & 100 & -4.89412   & 2.22541x$10^{-08}$ & -1.24304x$10^{-18}$ & 1.434x$10^{-29   }$ & 24.5921 & 4.30631x$10^{-12}$  \\ 
\rowcolor{gray!25} 31 & 20  & 0.01 & 795  & 50  & -2.25582   & 1.78805x$10^{-08}$ & -4.21493x$10^{-19}$ & -7.90319x$10^{-30}$ & 14.2263 & 4.19365x$10^{-12}$  \\ 
\rowcolor{white}   31 & 20  & 0.05 & 795  & 100 & -1.29916   & 4.11449x$10^{-09}$ & -3.06263x$10^{-20}$ & -2.27231x$10^{-31}$ & 5.16705 & 2.97909x$10^{-12}$  \\ 
\rowcolor{white}   31 & 20  & 0.05 & 795  & 50  & -0.378079  & 3.38521x$10^{-09}$ & -4.7016x$10^{-21 }$ & -2.66324x$10^{-31}$ & 2.80303 & 2.1794x$10^{-12 }$  \\ 
\rowcolor{gray!25} 31 & 350 & 0.01 & 795  & 100 & -0.36635   & 7.35375x$10^{-09}$ & -2.34247x$10^{-19}$ & 2.54538x$10^{-30 }$ & 18.9469 & 4.299x$10^{-12  }$  \\ 
\rowcolor{gray!25} 31 & 350 & 0.01 & 795  & 50  & -4.38687   & 8.44409x$10^{-09}$ & -2.43331x$10^{-19}$ & 2.3537x$10^{-30  }$ & 13.5755 & 4.88896x$10^{-12}$  \\ 
\rowcolor{white}   31 & 350 & 0.05 & 795  & 100 & -0.170798  & 1.38079x$10^{-09}$ & -3.2094x$10^{-21 }$ & -7.40306x$10^{-32}$ & 3.37071 & 1.43301x$10^{-12}$  \\ 
\rowcolor{white}   31 & 350 & 0.05 & 795  & 50  & -0.0616379 & 1.33379x$10^{-09}$ & -5.9227x$10^{-22 }$ & -6.15641x$10^{-32}$ & 2.23016 & 1.59918x$10^{-12}$  \\ 
\rowcolor{gray!25} 31 & 20  & 0.01 & 1200 & 100 & -2.53305   & 1.00688x$10^{-08}$ & -3.35282x$10^{-19}$ & 1.5248x$10^{-30  }$ & 22.3742 & 2.89434x$10^{-12}$  \\ 
\rowcolor{gray!25} 31 & 20  & 0.01 & 1200 & 50  & -1.04769   & 8.02545x$10^{-09}$ & -8.83194x$10^{-20}$ & -1.90744x$10^{-30}$ & 11.9071 & 3.0798x$10^{-12 }$  \\ 
\rowcolor{white}   31 & 20  & 0.05 & 1200 & 100 & -0.418796  & 1.83795x$10^{-09}$ & -3.60286x$10^{-21}$ & -1.38473x$10^{-31}$ & 4.50465 & 8.64053x$10^{-13}$  \\ 
\rowcolor{white}   31 & 20  & 0.05 & 1200 & 50  & -0.14429   & 1.53393x$10^{-09}$ & -2.17417x$10^{-22}$ & -5.46919x$10^{-32}$ & 2.22159 & 3.49267x$10^{-13}$  \\

    \end{tabular}
\end{table}

\newpage

\section*{Tutorial for Extending Kinematic DW Model}
\label{sec:tutorial}

The kinematic DW model may be extended to alternative parameter corners, input stimuli, or data; the following sections include scripts to automatically extract model fitting parameters to match new simulation or experimental data.

\subsection{Collecting Simulation Data} 

The Mumax3 DW script in the following \hyperref[sec:mumax]{section} may be altered to run simulations with different micromagnetic parameters or input stimuli.  For a fixed parameter corner, multiple simulations should be run with varying stimuli intensity.  The table.txt files generated by mumax contain all the data needed to calculate model fitting parameters.  

\subsection{Collecting Experimental Data}

Experiments should comprise a single uniform stimulus followed by a period without stimulus.  The timescale of both portions of the experiment should be sufficiently long for the DW to accelerate or decelerate to its final velocity, and multiple trials should be performed on the same material sample, varying only the strength of the input stimuli.  Each trial will have its own table and tables should be stored in folders according to the name convention specified in the following subsection.

Experimental data must match the Mumax3 table.txt format in order for the Matlab scripts to properly extract model fitting parameters -- that is in CSV format, structured according to the following table:

\begin{table}[h]
    \centering
    \begin{tabular}{|c|c|c|c|c|c|c|c|}
    \hline
         Time (s) & Unused & Unused & Unused & Norm. Mag. @ 0 nm & Norm. Mag. @ 1 nm & ... & Norm. Mag. @ (N) nm \\
    \hline
    0e-9 &  &  &  & 1 & 1 & ... & -1\\
         \hline
         1e-9 &  &  &  & 1 & 1 & ... & -1\\
         \hline
         $\vdots$ &  &  &  & $\vdots$ & $\vdots$ & $\ddots$ & $\vdots$ \\
         \hline
         (T)e-9 &  &  &  & 1 & 0.97 & ... & -1\\
    \hline
    \end{tabular}
    \label{tab:my_label}
\end{table}

The time step size may be arbitrary.  At each time step, magnetization of the sample should be measured at equally spaced locations along the length of the nanowire. For perpendicularly (in-plane) magnetized samples, the out-of-plane (track-length-wise) component of the magnetization should be written to the table, normalized to the saturation magnetization of the material. \textit{extract\_DW\_motion.m} will implicitly assume a position step size of 1 nm, matching the Mumax3 template; line 212  of \textit{extract\_DW\_motion.m} (positionStep = ...) should be changed to match the experimental spacing between samples.

\subsection{Extracting Model Parameters}

There are three Matlab scripts provided to automatically extract the kinematic DW model fitting parameters from micromagnetic simulation or experimental data.  The scripts are briefly described here and provided in a later section:

\subsubsection{extract\_DW\_motion.m}
This script can read in Mumax3 or experimental data formatted according the the previous subsection.  The script computes and saves (in .mat format) the DW position and velocity at each time step.

\subsubsection{DW\_analysis.m} 
All .mat files produced by \textit{extract\_DW\_motion.m} should be collected into folders, one folder per material parameter corner.  Each parameter corner folder should contain .mat files from several trials with varied stimulation strength. \textit{DW\_analysis.m} is run over an entire parameter corner folder and calculates the DW drift distance, velocity time constants, and maximum velocity for each trial, saving them into a single .mat file to be collected by \textit{fit\_model\_constants.m}.

\subsubsection{fit\_model\_constants.m}
All output .mat files from \textit{DW\_analysis.m} should be collected into a single folder.  \textit{fit\_model\_constants.m} extracts the kinematic model fitting coefficients for each parameter corner and compiles comprehensive lookup tables indexed by material parameter.

\subsection{Filename Conventions}

Parameter, stimuli, and timing information of the Mumax3 scripts are communicated in the file and folder names.  It is therefore important to follow the naming convention.  Mumax3 script filenames should be similar to the following:

\noindent\verb|DWSim_V=centerWall_Geom=1_Aex=11e-12_Ku=4.05e+5_A=0.01_Msat=7.95e+5_ ...|
\verb|u0Hke=NaN_DMI=NaN_J=4.0e+09_RT=100e-9_W=100e-9.mx3|

\noindent which is broken into a number of tokens, separated by single underscore characters.  Ordering of tokens does not matter.

The tokens "J" and "RT" correspond with "current density" (the stimulation strength) and "run time" (the duration stimulus is applied) and are mainly used by \textit{DW\_analysis.m} to extract the relationships between DW velocity and current density.

The other tokens represent the micromagnetic material parameters for that trial suite and are only used by \textit{fit\_model\_constants.m} to write the model parameter lookup tables which are indexed according to the various parameter corners.  Should parameters be added or removed from the filenames, the tracked parameter list in \textit{fit\_model\_constants.m} should be changed accordingly, in the sections beginning on line 60 and line 156.

\newpage

\section*{Mumax3 DW Script}
\label{sec:mumax}

\begin{verbatim}
//"DWSim_V=centerWall_Geom=1_Aex=11e-12_Ku=4.05e+5_A=0.01_Msat=7.95e+5_ ...
    u0Hke=NaN_DMI=NaN_J=4.0e+09_RT=100e-9_W=100e-9.mx3"

// sample mumax3 script for simple constant current pulse input
// The mumax script naming convention should be followed since extract_DW_motion.m 
//  uses the naming convention to associate the DW position with the micromagnetic 
//  parameters.

// 500nm x 100nm x 1.2nm track
setgridsize(500,63,1)
setcellsize(1e-9,1.6e-9,1.2e-9)

a := cuboid(500e-9, 100e-9, 1.2e-9)

defregion(0,a)

// set parameters
Aex.setregion(0, 11e-12)
alpha.setregion(0, 0.01)
//Xi.setregion(0, 0.01)
Msat.setregion(0, 7.95e+5)
Ku1.setregion(0, 4.05e+5)
anisU.setRegion(0, vector(0, 0, 1))
Dind.setRegion(0, 0.25e-3)

//m.loadfile("m_starter.ovf")
m = TwoDomain(0,0,1,  0,1,0,  0,0,-1)

//Converting SHE into Mumax3 STT
lambda = 1
Pol = 0.05 // to set thetaSH 0.05
epsilonprime = 0

//SHE related parameters(Limited to Domain Wall track)
sizeZ:= 1.2e-9
hcut := 1.0546e-34
thetaSH := 0.05
e := 1.6e-19
Ms:= 1.5/(pi*4e-7)
mull:=2
fixedlayer.setregion(0, vector(0,1,0) )


relax()

// keep DW centered in simulation window by shifting simulation window
ext_centerWall(2)

//autosnapshot(m, 10e-12)

// save magnetization state and simulation window shift amount to table, ...
    use to calculate position
for i := 0; i <= 499; i += 1 {
tableadd(Crop(m.comp(2), i, (i+1),30,32,0,1))
}
tableadd(ext_dwpos)
tableadd(ext_dwspeed)
tableautosave(1e-11)

//save ovfs if needed

//autosave(m,10e-11)

// simulate simple constant current pulse
Jelec:= 4.0e+09
j.setregion(0, vector(0, 0,Jelec))
B_ext.setregion(0, vector(0,mull*hcut*thetaSH*Jelec/(2*e*Ms*sizeZ),0))

run(100e-9)

Jelec = 0
j.setregion(0, vector(0, 0,Jelec))
B_ext.setregion(0, vector(0, mull*hcut*thetaSH*Jelec/(2*e*Ms*sizeZ),0))

run(100e-9)
\end{verbatim}

\newpage

\section*{Fitting Parameter Extraction}
\label{sec:matlab}

\subsection*{extract\_DW\_motion.m}

\begin{verbatim}
close all
clear

%read in mumax table files from specified folder, output excel and .mat files with ...
position and velocity extracted

% Specify the folder where the files live.
myFolder = '.\Simulations';

% Set to 1 to save data excel sheets along with the default .mat format
% Requires xlswritefig package from https://mathworks.com/matlabcentral/fileexchange/...
24424-xlswritefig
write_xlsx = 0;

% Check to make sure that folder actually exists.  Warn user if it doesn't.
if ~isfolder(myFolder)
    errorMessage = sprintf('Error: The following folder does not exist:\n%s\n ...
    Please specify a new folder.', myFolder);
    uiwait(warndlg(errorMessage));
    myFolder = uigetdir(); % Ask for a new one.
    if myFolder == 0
         % User clicked Cancel
         return;
    end
end

% Get a list of all files in the folder with the desired file name pattern.
filePattern = fullfile(myFolder, '*.out'); % Change to whatever pattern you need.
theFiles = dir(filePattern); %pull the files with this pattern from directory

%for loop to take action on all the files one at a time
for k = 1 : length(theFiles)
    figure(k); %create figure
    
    %get the name of the files & tell user which your reading each loop
    baseFileName = theFiles(k).name;
    fullFileName = fullfile(theFiles(k).folder, baseFileName);
    fprintf(1, 'Now reading %s\n', fullFileName);
    
    % Now do whatever you want with this file name (i.e. read it in)
    
    %file name example for this specific project:
    %DWSim_V=centerWall_Geom=1_Aex=31e-12_Ku=5.36e+5_A=0.05_Msat=7.95e+5_ ...
    u0Hke=NaN_DMI=NaN_J=2.8e+10_RT=20e-9_W=50e-9

    %fileNameParsed is a string array containing
    %each of the parameters for the mumax3 code
    fileNameParsed = split(fullFileName, "_");
    
    %create an excel file and write parameters to it
    if write_xlsx == 1
        fullNameExcel = [fullFileName '.xlsx'];
        writecell(fileNameParsed,fullNameExcel);
    end
    
    %need to read in the table.txt file from this file
    %this is what we will be performing the original matlab code on
    tableFile = [fullFileName '\table.txt'];
    
    %extract the data we will be analyzing
    dataAnalyzed = dlmread(tableFile, '',1,0);
    
    %write data to excel spreadsheet
    %dataTXT = fileread(tableFile);
    %dataXLS = 'table.xlsx';
    %T = readtable(tableFile);
    %writetable(T, dataXLS);
    
    %copied OneNeuron.m here to perform
    %the tests on the data acquired
   
    %looking for number of rows & columns
    %or [rows, columns] = size(dataAnalyzed);
    rows = size(dataAnalyzed,1);
    columns = size(dataAnalyzed,2);
    
    %shift the data in order to find position/velocity values
    dataShift = [ones(rows,1) dataAnalyzed(:,5:columns-3)]; % original data ...
    shifted one unit to the right
    dDiff = dataShift-dataAnalyzed(:,5:columns-2); % Think of this as a discrete ...
    derivative of d (dd/dx)
    x = 5:columns-2;
    
    [~, filename_noext, ~] = fileparts(baseFileName);
    
    %get what J (current density) is for DW is
    jVal = regexp(filename_noext, '_J=([^_]+)', 'tokens');
    jVal = jVal{1}{1};
    jDW = str2double(jVal);
    
    %get runtime for this DW    
    rtVal = regexp(filename_noext, '_RT=([^_]+)', 'tokens');
    rtVal = rtVal{1}{1};
    rtDW = str2double(rtVal);
    
    %reset the current
    Current = zeros(1,rows);
    
    %set the current based on where you are in runtime
    for i=1:rows
        if(dataAnalyzed(i,1) <= rtDW)
            Current(i) = jDW;
        else
            Current(i) = 0;
        end
    end

    %reinitialize the domain wall's position
    dwPosition = zeros(1,rows);

    for i=1:rows
        dwPosition(i)= sum(dDiff(i,:).*x) / sum(dDiff(i,:)); %integral of dd/dx ...
        * x * dx divided by integral of dd/dx dx
    end
    dwPositionShift = dwPosition + dataAnalyzed(:,columns-1)'*1e9 - dwPosition(1);

    dwPosition1 = dwPosition( floor((rows+1)/2) );        %nm
    dwPosition0 = dwPosition(1);                          %nm
    calculatedSlope = (dwPosition1 - dwPosition0) * 1e-9/(rtDW); %m/s

    %mute figures
    set(0, 'DefaultFigureVisible', 'off');
    
    %create the first figure for domain wall position vs time on y axis
    %and current vs time for yy axis
    figure(1);
    plot(dataAnalyzed(:,1)'*1e9,dwPositionShift(:),'b');
    hold on;
    ylabel ('Domain wall position (nm)');
    xlabel ('Time (ns)');
    yyaxis right;
    ylabel ('Current Density (10^1^2 Am^-^2)','Color','r' );
    p = plot(dataAnalyzed(:,1)'*1e9,Current/1e12,'r');
    
    %second page should be the plot from the data gathered above for both current
    %and position
    if write_xlsx == 1
        xlswritefig(p,fullNameExcel,'Sheet2','A1');
    end 
    hold off;
    
    %print the plot points on sheet three that coincide with the figures
    fig = findobj(figure(1), 'Type', 'Line' );
    
    x = get(fig, 'Xdata');
    y = get(fig, 'Ydata');
    
    out1 = 'Current Density (10^1^2 Am^-^2) By Time (ns)';
    out2 = 'Domain wall position (nm) By Time (ns)';
    
    if write_xlsx == 1
        try
            xlswrite(fullNameExcel,cellstr(out1),'Sheet3','A1');
            xlswrite(fullNameExcel,x{1},'Sheet3','A2');
            xlswrite(fullNameExcel,y{1},'Sheet3','A3');
        catch
            disp("couldn't write to excel");
        end

        try
            xlswrite(fullNameExcel,cellstr(out2),'Sheet3','A5');
            xlswrite(fullNameExcel,x{2},'Sheet3','A6');
            xlswrite(fullNameExcel,y{2},'Sheet3','A7');
        catch
            disp("couldn't write to excel");
        end
    end
    
    %next need to calculate the velocity
    figure(2); %create new figure
    time = dataAnalyzed(:,1);
    n = 2;
    dwPositionShift = dwPositionShift';
    dwNext = circshift(dwPositionShift, [n 0]);
    dwNext(1:n)=0;
    
    timeChange = 2;
    timeNext = circshift(time, [timeChange 0]);
    timeNext(1:timeChange)=0;
    
    delta = (dwPositionShift - dwNext)./(time - timeNext);
    
    %replot the position vs time on y axis & plot velocity vs time on
    %yy axis
    plot(dataAnalyzed(:,1)'*1e9,dwPositionShift(:),'b');
    hold on;
    
    %plot change in position/time to get velocity
    ylabel ('Domain wall position (nm)');
    xlabel ('Time (ns)');
    yyaxis right;
    ylabel ('Domain wall velocity (m/s)','Color','r' ); % change to red
    grid on;
    
    v = plot(time*1e9,delta/1e9,'r');
    
    %add velocity plot to second page
    %xlswritefig(v,fullNameExcel,'Sheet2','M1');

    fig2 = findobj(figure(2), 'Type', 'Line' );
    
    x2 = get(fig2, 'Xdata');
    y2 = get(fig2, 'Ydata');
    
    out3 = 'Domain wall velocity (nm) By Time (ns)';
    
    if write_xlsx == 1
        try
            xlswrite(fullNameExcel,cellstr(out3),'Sheet3','A9');
            xlswrite(fullNameExcel,x2{1},'Sheet3','A10');
            xlswrite(fullNameExcel,y2{1},'Sheet3','A11');
        catch
            disp("couldn't write to excel");
        end
    end
    
    % save time, position, and velocity to .mat file
    time = time';
    positionStep = 1e-9;
    dwPosition = dwPositionShift'*positionStep;
    dwVelocity = (delta*positionStep)';
    save(fullfile(theFiles(k).folder, strcat(baseFileName, '.mat')), 'time', ...
    'dwPosition', 'dwVelocity');
    
    %next subtract these
    clf(figure(1));
    clf(figure(2));
    
end
\end{verbatim}

\subsection*{DW\_analysis.m}

\begin{verbatim}
close all;
clear;

% Calculates various quantities (drift distance, time constant of velocity, ...
max velocity) for each current density
% Creates a table compiling calculated quantities for each current density within ...
a parameter corner
% Should be run after extract_DW_motion, using the output .mat files

% Specify the folder where the mat files live - folder contains all mat
% files for particular corner over multiple current densities
% example folder format: ".\Simulations\dataAnalysis\W=50e-9\Aex=11e-12\ ...
Ku=1.11e+6_A=0.05_Msat=1.2e+6"
folderPattern = '.\Simulations\dataAnalysis\W=50e-9\Aex=11e-12\Ku=*'; 

% if there are multiple folders matching pattern, each corresponding to a
% different parameter corner, loop through folders
folders = dir(folderPattern);
for n = 1: length(folders)
    myFolder = fullfile(folders(n).folder, folders(n).name);
    disp(myFolder);
    
    % Check to make sure that folder actually exists.  Warn user if it doesn't.
    if ~isfolder(myFolder)
        errorMessage = sprintf('Error: The following folder does not exist:\n%s\n ...
        Please specify a new folder.', myFolder);
        uiwait(warndlg(errorMessage));
        myFolder = uigetdir(); % Ask for a new one.
        if myFolder == 0
             % User clicked Cancel
             return;
        end
    end

    % Get a list of all files in the folder with the desired file name pattern.
    fileType = '*out.mat'; %xlsx or mat
    filePattern = fullfile(myFolder, fileType); % Change to whatever pattern you ...
    need.
    theFiles = dir(filePattern); %pull the files with this pattern from directory

    jVals = zeros(1,length(theFiles));
    maxVel = zeros(1,length(theFiles));
    timeConstant = zeros(1,length(theFiles));
    driftDist = zeros(1,length(theFiles));
    velJump = zeros(1,length(theFiles));

    % loop through each simulation .mat file
    for k = 1 : length(theFiles)

        %get the name of the files & tell user which your reading each loop
        baseFileName = theFiles(k).name;
        fullFileName = fullfile(theFiles(k).folder, baseFileName);
        fprintf(1, 'Now reading %s\n', fullFileName);

        %fileNameParsed is a string array containing
        %each of the parameters for the mumax3 code
        fileNameParsed = split(baseFileName, "_");

        %read in data from either excel or mat file
        if(strcmp(fileType, '*.xlsx'))
            %get time, position, and velocity from excel sheet
            time = readmatrix(fullFileName,'Sheet',3,'Range','2:2');
            dwPosition = readmatrix(fullFileName,'Sheet',3,'Range','7:7');
            dwVelocity = readmatrix(fullFileName,'Sheet',3,'Range','11:11');
        elseif(strcmp(fileType, '*out.mat'))
            load(fullFileName);
        end

        [~, filename_noext, ~] = fileparts(baseFileName);
    
        %get what J (current density) is for DW is
        jVal = regexp(filename_noext, '_J=([^_]+)', 'tokens');
        jVal = jVal{1}{1};
        jDW = str2double(jVal)
    
        %get runtime for this DW
        rtVal = regexp(filename_noext, '_RT=([^_]+)', 'tokens');
        rtVal = rtVal{1}{1};
        rtDW = str2double(rtVal)

        %reset the current
        Current = zeros(1,length(time));

        %set the current based on where you are in runtime
        current_end = -1;
        for i=1:length(time)
            if(time(i) <= rtDW)
                Current(i) = jDW;
            else
                if(current_end == -1)
                    current_end = i;
                end
                Current(i) = 0;
            end
        end

        % smooth velocity
        smoothVel = smoothdata(dwVelocity,'gaussian',150);

        %plot position/velocity figure
        
        figure(1); %create new figure

        %plot the position vs time on y axis & plot velocity vs time on
        yyaxis left;
        plot(time*1e9,dwPosition*1e9,'b','DisplayName','DW position');
        hold on;
        ylabel ('Domain wall position (nm)');
        xlabel ('Time (ns)');

        yyaxis right;
        ylabel ('Domain wall velocity (m/s)','Color','r' );
        set(gca,'ycolor','r');
        plot(time*1e9,smoothVel,'Color','r','DisplayName','DW velocity');
        legend()


        %save plot as png
        try
            saveas(figure(1), strcat(fullFileName, 'smooth.png'));
        catch
            disp('Couldn''t save figure');
        end
        %}

        %plot with current density
        %{
        figure(2);
        plot(time*1e9,dwPosition*1e9,'b','DisplayName','DW position');
        hold on;
        ylabel ('Domain wall position (nm)');
        xlabel ('Time (ns)');
        yyaxis right;
        ylabel ('Current Density (10^1^0 Am^-^2)','Color','r' );
        p = plot(time*1e9,Current/1e10,'r','DisplayName','Current Density');
        ylim([-Current(1)/1e10*0.1 Current(1)/1e10*1.1 ]);
        set(gca,'ycolor','r');
        legend()
        %saveas(figure(2), strcat(fullFileName,'_current', '.png'));
        %}
        
        disp(['Final position: ' num2str(dwPosition(end))]);

        %save current, maxVel, velocity time constant, driftDist, velJump, compile
        %in dataTable to fit model constants to simulation
        jVals(k) = jDW;
        maxVel(k) = max(abs(smoothVel(find(time>rtDW-1,1):current_end-1)));                         % max velocity reached (velocity right before current off)
        timeConstant(k) = time(find(abs(smoothVel-smoothVel(current_end-1)) ...
        <maxVel(k)/exp(1), 1));  % time when velocity reaches (1-1/e)*maxVel
        driftDist(k) = abs(dwPosition(end)-dwPosition(current_end));                                % distance traveled after current off
        
        % warn if DW still moving at end of simulation, need longer runtime
        if(abs(smoothVel(end))>0.01*maxVel(k))
            disp("DW not stopped")
        end

        % clear figures
        clf(figure(1));
        clf(figure(2));
    end

    %for naming, get string with Aex, Ku, A, Msat, W
    Aex_str = strcat('Aex=', extractBetween(baseFileName,'Aex=','_'), '_');
    Ku_str = strcat('Ku=', extractBetween(baseFileName,'Ku=','_'), '_');
    A_str = strcat('A=', extractBetween(baseFileName,'A=','_'), '_');
    Msat_str = strcat('Msat=', extractBetween(baseFileName,'Msat=','_'), '_');
    W_str = strcat('W=', extractBetween(baseFileName,'W=','.'));
    substr = char(strcat(Aex_str, Ku_str, A_str, Msat_str, W_str));
    disp(substr);

    % compile quantities into single table, sort table by current density
    dataTable = [jVals; maxVel; timeConstant; driftDist];
    dataTable = sortrows(dataTable')';

    % save table to same folder
    save(fullfile(theFiles(1).folder, strcat('dataTable_',substr,'.mat')), ...
    "dataTable");
end

\end{verbatim}

\subsection*{fit\_model\_constants.m}

\begin{verbatim}
close all;
clear;

% Code for finding model constants from mumax simulation fitting
% Use after running DW_analysis.m for each corner, and compiling all corners' ...
.mat file outputs into single folder
% Create and save lookup tables for model constants

% Specify the folder where the mat files live.
myFolder = 'Simulations\dataAnalysis\lookupTableData';

% Make sure the folder exists
if ~isfolder(myFolder)
    errorMessage = sprintf('Error: The following folder does not exist:\n%s\n ...
    Please specify a new folder.', myFolder);
    uiwait(warndlg(errorMessage));
    myFolder = uigetdir(); % Ask for a new one.
    if myFolder == 0
         % User clicked Cancel
         return;
    end
end

% Get a list of all files in the folder with the desired file name pattern.
fileType = "*.mat"; 
filePattern = fullfile(myFolder, fileType); 
theFiles = dir(filePattern); % pull all .mat files from directory

%create interp tables for constants of model fitting
interp_maxVel_c0 = [];
interp_maxVel_c1 = [];
interp_maxVel_c2 = [];
interp_maxVel_c3 = [];
interp_drift_const = [];
interp_d2 = [];

% list of param corner to keep track of order constants were saved to above tables
param_list = [];


%loop through all param corner .mat files
for k = 1 : length(theFiles)
    
    %load matrix (dataTable) with 4 rows: current, and corresponding ...
    maxVel/timeConstant/driftDist
    baseFileName = theFiles(k).name;
    fullFileName = fullfile(theFiles(k).folder, baseFileName);
    load(fullFileName);

    % use ind as upper index, to prevent fitting to data points with J>4e10
    ind = find(dataTable(1,:)>=4e10, 1); 
    if(isempty(ind))
        ind = length(dataTable(1,:));
    end

    % create seperate lists for quantities read in from .mat file
    J = dataTable(1,1:ind);
    maxVel = dataTable(2,1:ind);
    timeConstant = dataTable(3,1:ind)*1e9;
    driftDist = dataTable(4,1:ind)*1e9;
    disp(dataTable);
    
    %get parameter values for current parameter set
    [~, filename_noext, ~] = fileparts(baseFileName);
    
    AexVal = regexp(filename_noext, '_Aex=([^_]+)', 'tokens');
    AexVal = AexVal{1}{1};
    Aex_s = str2double(AexVal) * 1e12; %pJ/m
    
    KuVal = regexp(filename_noext, '_Ku=([^_]+)', 'tokens');
    KuVal = KuVal{1}{1};
    Ku_s = str2double(KuVal); %J/m^3
    
    AVal = regexp(filename_noext, '_A=([^_]+)', 'tokens');
    AVal = AVal{1}{1};
    A_s = str2double(AVal); %Unitless
    
    MsatVal = regexp(filename_noext, '_Msat=([^_]+)', 'tokens');
    MsatVal = MsatVal{1}{1};
    Msat_s = str2double(MsatVal); %A/m
    
    WVal = regexp(filename_noext, '_W=([^_]+)', 'tokens');
    WVal = WVal{1}{1};
    W_s = str2double(WVal) * 1e9; %nm
    
    
    % Using B anis in lookup table instead of Ku
    if(Ku_s == 1.11e6 || Ku_s == 5.36e5)
        B_anis_s = 350;
    elseif(Ku_s == 9.17e5 ||Ku_s == 4.05e5)
        B_anis_s = 20;
    else
        B_anis_s = (Ku_s/(0.5*Msat_s) - (4*pi*1e-7)*Msat_s)*1000;
    end

    % keep order of param coords to match interp_ matricies for later interp
    current_param = [Aex_s,B_anis_s,A_s,Msat_s,W_s];
    param_list = cat(1,param_list,current_param);
    
    
    %only include current range where max velocity increases with current
    if(max(maxVel) ~= maxVel(end))
        maxJind = find(maxVel==max(maxVel),1);
        
        J = J(1:maxJind);
        maxVel = maxVel(1:maxJind);
        disp(maxVel);
        timeConstant = timeConstant(1:maxJind);
        driftDist = driftDist(1:maxJind);

        disp(strcat("Changing current maximum to ",sprintf('%0.2e',J(end))," for ...
        ",mat2str(current_param)));
    end
    
    %fit maxVel to eqn a*J^3 + bJ^2 + cJ + d, use Jq/Jq(1) to scale down, ...
    adjust coeffs later
    J_fit = J/J(1);
    %fit with weight to make lower values more important & reduce % err at ...
    low values
    w = maxVel.^-2;
    cubicModel = @(b,x) b(1).*x.^3 + b(2).*x.^2 + b(3).*x + b(4);
    start = [1; 1; 1; 1];
    cubicFit = fitnlm(J_fit,maxVel,cubicModel,start,'Weight',w);
    cubic_coeffs = cubicFit.Coefficients.Estimate;
    
    % adjust coeffs to work for unscaled J
    c3 = cubic_coeffs(1)/(J(1)^3);
    c2 = cubic_coeffs(2)/(J(1)^2);
    c1 = cubic_coeffs(3)/J(1);
    c0 = cubic_coeffs(4);
    
    %fit drift dist to linear model
    driftDistPlot = [maxVel; driftDist];
    %add weight to earlier (smaller valued) part of fit
    w = driftDistPlot(2,:).^-1;
    linModel = @(b,x) b(1).*x;
    start = 1;
    linFit = fitnlm(driftDistPlot(1,:),driftDistPlot(2,:),linModel,start,'Weight',w);
    drift_fit = linFit.Coefficients.Estimate;
    
    
    %time const has inverse proportional relationship to J
    d2 = J'\(timeConstant.^-1 - 1/drift_fit)';
    if(d2<0)
        d2 = 0;
    end
    
    % update lists with model constants for this corner
    interp_maxVel_c0 = cat(1,interp_maxVel_c0,c0);
    interp_maxVel_c1 = cat(1,interp_maxVel_c1,c1);
    interp_maxVel_c2 = cat(1,interp_maxVel_c2,c2);
    interp_maxVel_c3 = cat(1,interp_maxVel_c3,c3);
    interp_drift_const = cat(1,interp_drift_const,drift_fit);
    interp_d2 = cat(1,interp_d2,d2);
    
    
end

% write lookup tables to .tbl files
% one lookup table per model constant, lists all param corners with ...
corresponding model constant value

fid = fopen(fullfile(myFolder, 'lookup_maxVel_c0.tbl'),'w+');
fprintf(fid,'#Aex(*1e12), B_anis, A, Msat, W(nm), c0 \n');
formatSpec = '%g %g %g %g %g %g\n';
fprintf(fid,formatSpec,cat(2,param_list,interp_maxVel_c0)');
fclose(fid);

fid = fopen(fullfile(myFolder, 'lookup_maxVel_c1.tbl'),'w+');
fprintf(fid,'#Aex(*1e12), B_anis, A, Msat, W(nm), c1 \n');
formatSpec = '%g %g %g %g %g %g\n';
fprintf(fid,formatSpec,cat(2,param_list,interp_maxVel_c1)');
fclose(fid);

fid = fopen(fullfile(myFolder, 'lookup_maxVel_c2.tbl'),'w+');
fprintf(fid,'#Aex(*1e12), B_anis, A, Msat, W(nm), c2 \n');
formatSpec = '%g %g %g %g %g %g\n';
fprintf(fid,formatSpec,cat(2,param_list,interp_maxVel_c2)');
fclose(fid);

fid = fopen(fullfile(myFolder, 'lookup_maxVel_c3.tbl'),'w+');
fprintf(fid,'#Aex(*1e12), B_anis, A, Msat, W(nm), c3 \n');
formatSpec = '%g %g %g %g %g %g\n';
fprintf(fid,formatSpec,cat(2,param_list,interp_maxVel_c3)');
fclose(fid);

fid = fopen(fullfile(myFolder, 'lookup_drift_const.tbl'),'w+');
fprintf(fid,'#Aex(*1e12), B_anis, A, Msat, W(nm), drift_const \n');
formatSpec = '%g %g %g %g %g %g\n';
fprintf(fid,formatSpec,cat(2,param_list,interp_drift_const)');
fclose(fid);

fid = fopen(fullfile(myFolder, 'lookup_d2.tbl'),'w+');
fprintf(fid,'#Aex(*1e12), B_anis, A, Msat, W(nm), d2 \n');
formatSpec = '%g %g %g %g %g %g\n';
fprintf(fid,formatSpec,cat(2,param_list,interp_d2)');
fclose(fid);



\end{verbatim}


\bibliographystyle{IEEEtran}
\bibliography{main}
\label{sec:references}